\documentclass[11pt,preprint]{aastex}

\usepackage{amsmath} 

\def\beq{\begin{equation}} 
\def\eeq{\end{equation}} 
\def\beqar{\begin{eqnarray}} 
\def\eeqar{\end{eqnarray}} 

\def\pref#1{(\ref{#1})}
 
\def\pfrac#1#2{\left( \frac{#1}{#2} \right)} 
\def\avg#1{\langle #1 \rangle}

\def\msol{M_\odot}
 
\def\csfr{\dot{\rho}_\star} 
\def\csnr{{\cal R}_{\rm SN}} 

\def\Mpeak{M_{\rm peak}}
\def\snlf{\phi_{\rm snlf,x}}

\def\mag#1{#1^{\rm mag}}

\def\mlim{m_{\rm lim}^{\rm sn}} 
\def\oscan{\Delta \Omega_{\rm scan}}

\def\la{\mathrel{\mathpalette\fun <}}
\def\ga{\mathrel{\mathpalette\fun >}}
\def\fun#1#2{\lower3.6pt\vbox{\baselineskip0pt\lineskip.9pt
  \ialign{$\mathsurround=0pt#1\hfil##\hfil$\crcr#2\crcr\sim\crcr}}}

\begin{document} 



\title{Cosmic Core-Collapse Supernovae from Upcoming Sky Surveys} 
 
\author{Amy Lien and Brian D. Fields} 
 
\affil{Department of Astronomy, University of Illinois, Urbana, IL}

\begin{abstract} 
Large synoptic (repeated scan) imaging sky surveys  
are poised to observe enormous numbers of core-collapse supernovae. 
We quantify the discovery 
potential of such surveys, and apply our results to 
upcoming projects, including DES, Pan-STARRS, and LSST. 
The latter two will harvest core-collapse supernovae
in numbers
orders of magnitude greater than have ever been observed to date.
These surveys 
will map out the cosmic core-collapse supernova 
redshift distribution via direct {\em counting}, with 
very small statistical uncertainties 
out to a redshift depth which is a strong function of
the survey limiting magnitude.
This supernova redshift history encodes rich information about
cosmology, star formation, and supernova astrophysics and phenomenology;
the large statistics of the supernova sample will be crucial to
disentangle possible degeneracies among these issues.
For example, the cosmic supernova {\em rate}
can be measured to high precision 
out to $z \sim 0.5$ for all core-collapse types,
and out to redshift $z \sim 1$ for Type IIn events
if their intrinsic properties remain the same
as those measured locally.
A precision knowledge of the cosmic supernova rate
would remove the cosmological uncertainties 
in the study of the wealth of observable  
properties of the cosmic supernova populations and their evolution
with environment and redshift. 
Because of the tight link between supernovae and star formation,
synoptic sky surveys will also provide
precision measurements of the normalization and
$z \la 1$ history of cosmic star-formation rate
in a manner independent of and complementary to
than current data based on UV and other proxies
for massive star formation.
Furthermore, Type II supernovae can serve as distance
indicators and would independently cross-check
Type Ia distances measured in the same surveys.
Arguably the largest and least-controlled uncertainty in all of these
efforts comes from the poorly-understood evolution of
dust obscuration of supernovae in their host galaxies;
we outline a strategy to determine empirically the obscuration
properties by leveraging the large supernova samples over
a broad range of redshift.
We conclude with recommendations on how best to use (and to tailor) 
these galaxy surveys to fully extract unique new probes on the 
physics, astrophysics, and cosmology of core-collapse explosions.
\end{abstract} 

\noindent

\section{Introduction}

A new generation of deep, large-area, synoptic
(repeated-scan)
galaxy surveys is coming online and is
poised to revolutionize cosmology in particular and
astrophysics in general.
The scanning nature of these surveys will open
the way for a systematic study of the celestial sphere in the
time domain.
In particular, ongoing and planned surveys
are sensitive to the
transient cosmos on timescales from hours to
years, and to supernova flux limits down to 
$\mag{24}$
and sometimes fainter.
As we will see, these capabilities
will reap a huge harvest in cosmic supernovae
and will offer a new and direct probe of the
cosmic supernova history out to high redshifts.

In the past decade,
supernovae in nearby and distant galaxies have come to play
crucial role for cosmology, via the use
of Type Ia explosions
as ``standardizable''
candles \citep[e.g.,][]{phillips,riess96}.
These powerful beacons are detectable out to very
high redshift and thus reveal the cosmic expansion history
for much of the lifetime of the universe;
the stunning result has been the detection of
the acceleration of the Universe and the
inference that dark energy of some form dominates
the mass-energy content of the cosmos today
\citep[e.g.,][]{riess,perlmutter,astier,wood-vasey}.
The detection of large numbers of Type Ia supernovae over
a large redshift range, and their
use as cosmological probes, represents a major focus of future
galaxy surveys \citep[e.g.,][]{wang}.

While studies of supernova Type Ia (thermonuclear explosions)
justly
receive enormous attention due to their cosmological
importance, there has been relatively little
focus on the detection of 
the more numerous population of
core-collapse supernovae.
These explosions of massive stars 
show great diversity in their observed properties,
e.g.~including several varieties of
Type II events, but also Types Ib and Ic 
events. 
Despite their heterogeneous nature, some core-collapse events
may nonetheless provide standardized candles, via
their early lightcurves whose nature is set
by the physics of their expanding photospheres
\citep[][see below]{kk,baron,dh}.
Moreover,  core-collapse 
events are of great intrinsic importance for
cosmology, astrophysics, and particle physics.
These events play a crucial role 
in cosmic energy feedback
processes and thus
in the formation and evolution of
galaxies and of cosmological structure.

Synoptic surveys tuned for Type Ia events will also automatically
detect core-collapse supernovae. Indeed,
as survey coverage and depth increase,
they will, for the first time, image a large fraction of 
{\em all} unobscured
cosmic core-collapse supernovae out to moderate redshift.
These photometric detections of supernovae and their
light curves will shed new light on a wide variety of 
problems spanning cosmology, particle astrophysics, and
supernova studies.
Moreover, these data will ``come for free''
so long as surveys include core-collapse events in their analysis pipelines.

For example,
\citet{madau98} already pointed out the
link between the cosmic star formation history
and the cosmic supernova history, and 
showed that when integrated over all redshifts,
the all-sky supernova event
rate is enormous, $\simeq 5-15$ events/sec in their estimate.
Upcoming synoptic surveys will probe most or all of 
the sky at great depth, and thus are positioned to
observe a large fraction of these events.
Consequently, these surveys will reveal the history of cosmic
supernovae via directly {\em counting} their numbers
as a function of redshift.

Already, recent and ongoing surveys have
begun to detect  core-collapse 
supernovae.
However, to date, surveys have focused on
Type Ia events, and thus core-collapse discovery and observation 
has been a serendipitous or even accidental byproduct of SNIa searches.
As a result of these surveys, the supernova discovery
rate is accelerating, and
the current all-time, all-Type supernova
count is $\sim 5000$ since SN1006.\footnote{
\citet{cbat}; 
see also {\tt \url{http://www.cfa.harvard.edu/iau/lists/Supernovae.html}}
}
Thus core-collapse data is currently sparsely analyzed
and reported in an uneven manner.
This situation will drastically improve in the near future,
when the supernova count increase by large factors,
culminating in up to $\sim$ 100,000 core-collapse events
seen by LSST annually.
In this paper we therefore will anticipate this future,
rather than make extensive comparison with the present data
though we will make quantitative contact
with current results.

Our work draws upon
several key analyses.
The thorough and elegant work of 
\citet{dahlen} laid out the framework for
rates and observability of cosmic supernovae
of all types.  Their work assembled a large body
of supernova data and applied it to make rate
and discovery 
predictions for the wide variety of star formation histories
and normalizations viable at that time,
with a particular focus on forecasts
for very high redshift (out to $z \sim 5$) observable
by the infrared {\it James Webb Space Telescope}.
\citet{sullivan} estimated the rates for  supernovae
lensed by the matter distribution--particularly rich clusters--along
the line of sight; these objects 
further extend the reach of infrared searches, and
identified a possible supernova candidate from {\em Hubble Space Telescope} archival images of an intermediate-redshift cluster.
\citet{gms} made similar calculations of the infrared observability of 
supernovae, and identified additional events in archival data.
\citet{gal-yam}
and \citet{oda} 
presented forecasts for then-upcoming ground-based surveys.
These studies considered all cosmic
supernovae, but with a focus 
on Type Ia events,
specifically with an eye towards revealing the
Type Ia delay time as well as
a parameterized characterization of
the cosmic star formation history
based on Type Ia counts.
In addition, these first studies reasonably chose to
emphasize near-term
(i.e., now-completed or ongoing) relatively
modest surveys, or on future space-based missions 
such as SNAP,
with little to no study of the
impact of large synoptic surveys.  
Moreover, while these works included dust extinction effects
in host galaxies, but because of their focus on Type Ia events,
they did not
study the possibility of a redshift evolution
in extinction \citep{mannucci}.

We build on the important studies of
\citet{dahlen},
\citet{gms,gal-yam}, and \citet{oda} in several respects:
(1) we explore the promise of
synoptic surveys
and forecast the very large numbers of supernovae they will find;
(2) we focus on less-studied core-collapse events;
(3) we incorporate the (pessimistic) possibility of 
strong dust evolution of
\citet{mannucci} which is a dominant obstacle to observing
massive star death at high redshifts;
(4) we present a strategy for {\em empirically}
calibrating the obscuration properties across a broad
range of redshift by studying the evolution of
the supernova luminosity function;
and (5) we study the unique opportunities that become
available with the large supernova harvest of synoptic surveys;
in particular, we show how the cosmic supernova rate
can be recovered based on core-collapse counts,
without assumption as to its functional form.

Our goal in this paper is to explore the 
impact synopic surveys will make on core-collapse supernova
astrophysics and cosmology.
We summarize key upcoming surveys in \S\ref{sect:surveys}.
In \S\ref{sect:inputs}
we review expectations for the CSNR,
core-collapse supernova observables,
and the effect of cosmic dust and its evolution.
We combine these inputs in \S\ref{sect:forecasts}
where we forecast
the core-collapse supernova discovery potential
for upcoming surveys. 
We quantify in detail the strong dependence 
of the supernova harvest on the survey limiting magnitude,
which we find to be the key figure of merit
for supernova studies.
We discuss some of the supernova science payoff in
\S\ref{sect:discuss},
and conclude in \S\ref{sect:conclude}
with some recommendations for
synoptic surveys.

\section{Synoptic Surveys}
\label{sect:surveys}

Current and future sky surveys 
build on the pioneering approach of the Sloan Digital Sky 
Survey \citep[SDSS;][]{sdss}.  Following SDSS, these surveys will
produce high-quality digital photometric maps of large
regions of the celestial sphere.
The powerful innovation the new surveys 
to extend the original SDSS approach into the
time domain.  Each program will scan part of their
survey domain frequently, with revisit periods of days
and in some cases even hours, and maintain
this systematic effort throughout the survey's
multi-year operating lifespans.
The result will be unprecedented catalogs of transient 
phenomena over timescales from hours to years.
These surveys are thus
ideal for supernova discovery and matched to
supernova light curve evolution timescales;
the result will be a revolution in our observational
understanding of supernovae.

\begin{table}[ht]
\caption{\label{tab:vitalstats} Recent and Future Synoptic Sky Surveys}
\begin{tabular}{c|cccc}
\hline\hline
Survey & Scan Area & SN Depth & Scan & Expected \\
Name & $\oscan \ [\rm deg^2]$ & $r$-band $\mlim \ [\rm mag]$ 
& Region & Operation \\
\hline
SDSS-II & 300 & 21.5 & SDSS southern equatorial strip & 2005--2008\\
DES & 40 & 24.2 & South Galactic Cap & 2011--2016 \\
Pan-STARRS & 30000 & 23 & $\sim 75\%$ of the Hawai'ian sky & 2010--2020 \\
LSST & 20000 & 23--25 & southern hemisphere & 2014--2024\\
\hline\hline
\end{tabular}
\end{table}

The science harvest in the time domain depends on 
both the depth of the scans and their breadth across the
celestial sphere.
These scale with collecting area $A$ and sky coverage
$\Omega_{\rm survey}$,
respectively.  Consequently, the figure of merit
for scan power
is the {\'e}tendue $A \Omega_{\rm survey}$.
Forthcoming projects are designed to maximize
this quantity.

The viability of supernova discovery, typing, and followup
by large-scale synoptic surveys has now been tested by
the SDSS-II supernova search \citep{frieman}.
This program extended SDSS \citep{sdss}
into the time domain, scanning at a $\sim 5$ day
cadence, identifying and typing supernova candidates from photometric data in
real time,
and following up with spectroscopic confirmation.  This survey
will serve as a testbed for the larger future campaigns.
It is thus very important and encouraging that SDSS-II 
has reported the discovery of 403 confirmed supernovae 
in the first two seasons of operation
\citep{sako}.  The search algorithms and followup were focused on
Type Ia events, for which light curves and spectra
have been recovered over $0.05 < z < 0.35$;
human input was used for supernova typing, but automated
routines appear promising and will be essential for larger surveys. 
Follow-up spectroscopy \citep{zheng}
yields accurate supernova and host-galaxy redshifts
($\sigma_z^{\rm sn} \approx 0.005$ and $\sigma_z^{\rm gal} \approx 0.0005$);
host-galaxy contamination is found to be well-addressed by $\chi^2$ fitting
and a principal component analysis.

Table \ref{tab:vitalstats}
lists several major current and future synoptic surveys,
and gives the values or current estimates of their
performance characteristics.
The $\mlim$ values are derived from the survey $5\sigma$
detections for single visit exposures, which have been
corrected $\mag{1}$ shallower as noted above.
SDSS-II \citep{frieman} is recently completed, as discussed above;
we adopt an $r$-band limiting magnitude of
$\mag{21.5}$ (J. Frieman, private communication).
The Dark Energy Survey \citep[DES;][]{des}
will push down to
$\mlim \sim \mag{24.2}$ in $r$-band; as we will see below,
this will already enormously increase the supernova
harvest.
Finally, looking out farther into the next decade,
Pan-STARRS \citep{pan-starrs,pan-starrs-sne}
and then LSST \citep{lsst,tyson}
will introduce a huge leap in both sky coverage
and in depth.
These ambitious projects represent a culmination
of the synoptic survey approach, and we will make
a particular effort to examine their potential for
supernova science.

For our analysis, we will 
characterize each survey with 
four parameters
\begin{enumerate}

\item
the survey supernova depth, i.e., single exposure limiting magnitude $\mlim$
for supernova detection when used in scan mode; this is set by collecting area
(and monitoring time)

\item 
the total survey scanning sky coverage, i.e., solid angle $\oscan$

\item
the scan revisit time (``cadence'')
$\tau_{\rm visit}$

\item
the total monitoring time $\Delta t_{\rm obs}$, which (for a single cadence)
is proportional to the total number 
$\Delta t_{\rm obs}/\tau_{\rm visit}$ of visits

\end{enumerate}
For a fixed survey design and lifetime, 
these parameters are not independent, 
since exposure time comes at the expense of sky coverage and number of visits.

There are numerous challenges and complexities
in the process of extracting supernovae and their redshifts from surveys
\citep[and for sorting out their types; see, e.g.,][]{dg,poznanski,
kim,kunz,blondin,wang2007}.
\citet{tonry} gives thorough discussion of these issues with
emphasis on Type Ia events; see also \citet{dahlen}, \citet{gal-yam}
and \citet{oda}, and the SDSS-II papers
\citep{frieman,sako,zheng}.

Our simple survey parameterization cannot capture all of these
subtleties, not do we intend it to; rather, we wish our
treatment to provide a rough illustration of the 
surveys' potential for core-collapse detection and science.
Consequently, our parameter choices should
be viewed as typical effective values, which
may be different from (and weaker than) the 
raw survey specifications.

For example, supernova identification
and typing requires knowledge of the light curve.
Thus, one cannot only observe the supernova
at peak brightness, but also follow it after (and ideally
before).  
\citet{pan-starrs-sne} recommends 
following the supernova for least $\delta m = \mag{1}$ below
peak brightness; we will adopt this value as well. 
Thus, the effective supernova detection depth is
$\mlim = m_{\rm max} - \delta m$,
where $m_{\rm max}$ is the survey scan depth
(i.e., depth for a single exposure).

Note also that some upcoming surveys (such as DES) will only 
repeatedly scan a fraction of the sky which they map;
but only the {\em scanned} regions
will host the discovery of supernovae and other transients. 
Also, some surveys (e.g., Pan-STARRS and LSST)
envision multiple periodicities and associated limiting magnitudes;
for simplicity we will here chose conservative depths for the values
given in Table \ref{tab:vitalstats},
to be consistent with the advertised scanning sky coverage.
Thus one should bear in mind that in our analysis
we have chosen  the minimal parameterization one could
use, which give only a simplified and idealized
sketch of the real surveys.
Given this, and the ongoing planning of future
survey characteristics, our forecasts for 
the surveys' supernova results should be understood
as indicative of the order of magnitude expected,
but not as high-precision  predictions.

\section{Core-Collapse Supernovae in a Cosmic Context} 
\label{sect:inputs}

\subsection{The Cosmic Core-Collapse Supernova Rate:  Expectations} 

The total cosmic supernova rate (hereafter CSNR) 
\beq
\csnr[z(t_{\rm em})] \equiv   \frac{dN_{\rm SN}}{dV_{\rm com} \, dt_{\rm em}} 
\eeq
is the number of events per
comoving volume
per unit time $t_{\rm em}$ in the emission frame
(i.e., cosmic time dilation effects in the observer's $z=0$ frame
are not included).
The total rate, and the various differential rates below,
can of course be specialized to distinguish different groups
of supernovae classified by intrinsic type and/or dependence
on local environment.

The present data on high-redshift core-collapse 
supernovae are too poor to construct an accurate CSNR.
But the CSNR is
intimately related to cosmic {\em star-formation} rate 
$\dot{\rho}_\star  = dM_{\star}/dV_{\rm com} dt$
\citep{madau98}.
The connection is
\beq
\label{eq:snr-sfr}
\csnr = \frac{X_{\rm SN}}{\avg{m_{\rm SN}}} \dot{\rho}_\star
\eeq
where $X_{\rm SN}$ is the fraction, by mass, of
stars which become supernovae, and $\avg{m_{\rm SN}}$
is the average supernova progenitor mass
(see Appendix \ref{sect:app-csfr}).
A key  point is that due to the short core-collapse progenitor lifetimes
the two rates scale linearly, $\csnr \propto \dot{\rho}_\star$.
The constant of proportionality
depends on the initial mass function (IMF).
If the IMF changes with time (or environment) this complicates
the picture.  In producing quantitative estimates
we will follow most studies in assuming time-independent IMF.
Thus the supernova/star-formation rate proportionality 
is a constant fixed for all time, namely
$\csnr/\dot{\rho}_\star = 0.00915 \, \msol^{-1}$
(Appendix \ref{sect:app-csfr}).

Uncertainties in
the cosmic rates for both supernovae and star-formation
remain considerable.
As illustrated in detail by
\citep{strigari05},
the cosmic star-formation rate is 
known to rise sharply towards redshift $z \sim 1$. 
In this low-to-moderate redshift regime,
the {\em shape }of the rate versus redshift is fairly
well known, but as emphasized by \citet{hopkins}
the {\em normalization} remains uncertain to within a factor $\sim 2$.
At higher redshifts, the rate becomes even more uncertain,
largely due to the paucity of data and also to uncertainties in 
our knowledge of the degree of dust obscuration.
It is also worth noting that most studies to date
directly or indirectly use massive stars as
proxies for star formation.  Consequently, 
the rate for cosmic massive star formation--and for cosmic supernovae--is 
less uncertain and IMF-dependent than the total rate. 

\begin{figure}
\includegraphics[width=0.8\textwidth]{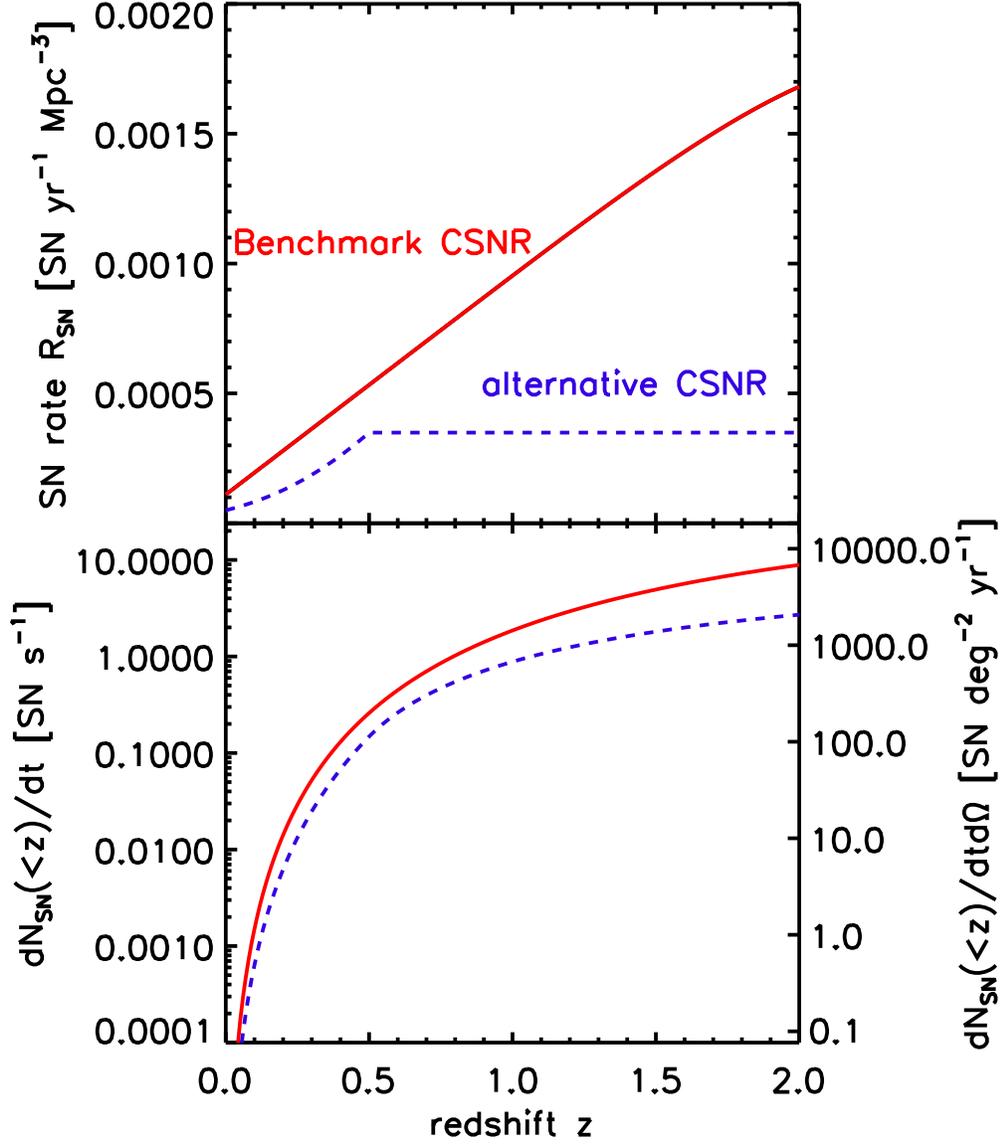}
\caption{
(a) {\em Top panel:} 
Possible cosmic core-collapse 
supernova rates as a function of redshift.
The solid curve is the result calculated based on the
\citet{cole} cosmic star-formation rate;
the broken curve is based on current supernova data
\citep{botticella};
see Appendix \ref{sect:app-csfr}. 
(b) {\em Bottom panel:}
The idealized, all-sky cumulative rate of all supernovae observed over
redshift 0 to $z$, for an observer with no faintness limit
and with no dust extinction anywhere along the line of sight.}
\label{fig:csnr}
\end{figure}

To illustrate the effects of these uncertainties on 
the synoptic survey supernova harvest,
we have adopted two possible CSNR forms.
These appear in 
Figure \ref{fig:csnr}, which
shows the expected supernovae rate assuming a perfect
environment (i.e. no dust extinction, etc). 
The solid curve in Figure \ref{fig:csnr}(a) is
the CSNR derived from 
the cosmic star-formation rate of \citet{cole} 
with parameters fitted by \citet{hopkins} (hereafter the ``benchmark'' CSNR).
This rate sharply rises to a peak at $z \sim 2.5$, followed 
a strong but less rapid declines out to high redshift.
To investigate the impact of the falloff from the peak, we also
show in the broken curve an alternate CSNR due to current observational data 
fitted by \citet{botticella} (hereafter the ``alternative'' CSNR).
This rate also rise to redshift
$z \sim 0.5$, though with a different slope;
we somewhat arbitrarily set the alternative rate to a constant at
$z > 0.5$ where the data are unclear; in any case we will find
that few events from this high-redshift regime will be accessible to the
all-sky surveys which are our focus.


Synoptic surveys will measure several
observables associated with cosmic supernovae:
their numbers and location, and some portion of their light curves
in different bands.  Spectroscopic redshifts of host galaxies
can also be determined
(when visible; see \S \ref{sect:followup}). 
Using the number counts and redshift indicators,
one can deduce an observed core-collapse rate, 
per unit redshift and per unit time and solid angle.
This observed rate distribution directly encodes the CSNR via 
\beq
\label{eq:idealrate}
\frac{dN_{SN}}{d\Omega dt_{\rm obs} dz} 
 =  \frac{dN_{SN}}{dV_{\rm com} dt_{\rm em}}
  \ \frac{dt_{\rm em}}{dt_{\rm obs}}
  \ \frac{dV_{\rm com}}{d\Omega dz}  
 = \csnr(z) \frac{r^2_{\rm com}}{1+z} \frac{dr_{\rm com}}{dz}
\eeq
where $V_{com}$ is the comoving volume and $r_{com}(z)$ is the comoving
distance out to redshift $z$.
The $1+z$ factor corrects for time dilation
via
$dt_{\rm obs} = (1+z) dt_{\rm em}$.

Figure \ref{fig:csnr}(b) shows the all-sky cumulative
frequency of cosmic supernovae for an observer
at $z=0$, i.e.,
\beq
\frac{dN_{\rm SN}}{dt}(<z)_{\rm all-sky}
 = 4\pi \int_0^z \frac{dN_{SN}}{d\Omega dt_{\rm obs} dz} dz^\prime
 = 4\pi \int_0^z \csnr(z^\prime) 
  \ \frac{r_{\rm com}^2}{1+z^\prime} \ \frac{dr_{\rm com}}{dz^\prime} \ dz^\prime 
\eeq
These curves give the total rate of observed cosmic supernova explosions
out to redshift $z$  for an idealized observer monitoring
the entire sky out to unlimited depth and without any 
dust obscuration anywhere along the line of sight.

All of these idealizations will fail, some of them drastically, for
real observational programs.  Nevertheless,
one cannot help but be tantalized by
the enormous explosion frequencies indicated in 
Figure \ref{fig:csnr}(b).  With our benchmark CSNR, out to
redshift $z=1$, something like $\sim 1$ supernova explodes
{\em per second} somewhere in the sky.
Out to redshift $z=2$, this rate increases
to $\sim 6$ events/sec.  Clearly, even with a small
detection efficiency, synoptic surveys 
are poised to discover core-collapse supernovae in numbers
far exceeding all supernovae in recorded history to date.

For numerical results in
Figure \ref{fig:csnr} and throughout this paper, we adopt a flat cosmology
with $\Omega_{\rm m} = 0.3$ and $\Omega_\Lambda = 0.7$.
For the Hubble constant
we adopt the value
$H_0 = 71 \ \rm km \, s^{-1} \, Mpc^{-1}$,
i.e., $h = 0.71$ where
$H_0 = 100 \ h \ \rm km \, s^{-1} \, Mpc^{-1}$.
These values are consistent with
recent determinations using WMAP
and large-scale structure \citep{wmap07}.

\subsection{The Effect of Dust Obscuration}
\label{sect:dust}

The enormous inventory of cosmic supernovae is not,
unfortunately, fully observable even for arbitrarily
deep surveys.
In a realistic environment there are several factors which
will hide the supernovae from us; dust extinction is one of the most
important, and probably the most uncertain. 
Core-collapse supernovae mostly explode within
regions of vigorous star formation
which are thus likely to be dusty environments. 
Consequently, we expect that some
core-collapse supernovae will be obscured to the point
where they are not detected in synoptic surveys.
The fraction of supernovae lost to dust obscuration,
and particularly the possible redshift dependence of this extinction,
represents a crucial systematic error which must be
addressed before one can use survey data to infer
information about supernova populations and their cosmic rates.

For the purposes of our present estimates of survey supernova yields, 
we follow the approach of
\citet{mannucci}. 
These authors characterize losses due to dust extinction and/or
reddening in the host galaxies via a fraction $\alpha_{\rm dust}(z)$ of
undetected events at each redshift.
This fraction could in principle differ for the various 
core-collapse types; for the present treatment we will assume it is
the same for all such events.  As core-collapse statistics become available
from surveys, this issue can and should be revisited; more on this below.
and in \S \ref{sect:dustfuture}.
The resulting fraction of {\em detected} supernovae
is thus the complement $f_{\rm dust}(z)  = 1 - \alpha_{\rm dust}(z)$, 
which measures the reduced supernova detection efficiency in
the presence of dust.
Expressed as an effective extinction $A$ for the
supernova population at $z$, we have
$A_{\rm eff}(z) = - 1.086 \ln f_{\rm dust}$.

\citet{mannucci} estimate the fraction of missing supernovae
by comparing the observed detections in the optical with those
in radio and near-IR.
They conclude that dust evolution is very strong;
this becomes a dominant limitation to the discovery of
core-collapse events at high redshift.
In the local 
universe,
\citet{mannucci}
find that the vast majority of the events occurring
in massive starbursts (luminous infrared galaxies)
are missed.  Because these galaxies
harbor only a small fraction of the local supernova population,
the overall
optically missing fraction at $z=0$
is estimated to be rather modest: $\alpha_{\rm dust} = 5-10\%$ .
If, however, 
high-redshift star formation occurs in starburst environments
\citep[i.e., luminous and ultra-luminous galaxies, which are
highly extincted; see e.g.][]{smail,hughes,perez-gonzalez,lefloch,choi}
then the
fraction of missing events rises sharply with redshift. 
Multiwavelength observations of light from pre-supernova
massive stars also supports the idea of increasing
dust obscuration at high redshift.
\citet{adelberger} find that 
ultraviolet light from massive stars in $z \sim 3$ galaxies
is mostly reprocessed by dust into thermal
submillimeter emission, so that the observable
galaxy luminosities have $L_{\rm sub-mm}/L_{\rm UV} \sim 1-100$.

\citet{mannucci} estimate the portion of supernovae
which will be ``catastrophic losses'' to severe extinction,
and propose that the missing fraction can be
described by a linear relation 
$\alpha_{\rm dust}(z)=0.05+0.28z$ for the core-collapse supernovae for
redshift z$<$2.
Thus,
the fraction of the supernovae which remain optically
detectable is
$f_{\rm dust}(z)  = 1-\alpha_{\rm dust}(z) = 0.95 - 0.28 z$
for $z<2$. 
At higher redshift, 
\citet{chen}
and \citet{gnedin}
argue that $f_{\rm dust}$ is small;
they find limits consistent with 
$f_{\rm dust} = 0.02$ for these redshifts.

We will smoothly match these two estimates,
and adopt a 
fraction of the supernovae which can be
detected after dust extinction of
\beq
\label{eq:dust}
f_{\rm dust}(z)
  = \left\{
   \begin{array}{cc}
      0.95 - 0.28 z  \ , & z < 3.3\\
      0.02  \ , & z \ge 3.3
   \end{array}
  \right.
\eeq
For these values of $f_{\rm dust}$, the effective extinction
varies from $A_{\rm eff} = \mag{0.056}$ at $z=0$
to $A_{\rm eff} = \mag{4.25}$ at $z \ge 3.3$.
In practice, we will find that cosmological dimming of
supernovae beyond $z \sim 1$
is itself so large that surveys up to and including
LSST will see relatively few events,
so that the details of the adopted dust model in this regime
will not affect our conclusions.

The strong redshift evolution of dust obscuration in the
empirical \citet{mannucci} model deserves comment.
From a physical point of view, the rise in dust losses $\alpha(z)$ towards
high redshift implies that at earlier times, 
the birth environments of supernovae are significantly more
enshrouded than those now.
This interesting result itself deserves a deeper
elucidation, one which will likely be easier to formulate and 
test in the presence of survey supernova data.  
From more practical point of view, 
our adoption of a model wherein dust losses grow rapidly
with $z$ should yield conservative (or at least not optimistic)
predictions for the supernova harvest at large redshifts.
That is, if it turns out that host galaxy effects do not
change rapidly with cosmic time so that the
efficiency of supernova detection remains close to the
high local value, then our rate predictions at $z \sim 1$ would 
be boosted by a factor of $\sim 1.5$.

Note also that $f_{\rm dust}$ as \citet{mannucci} 
and we have defined it
characterizes the observable portion of
the {\em ensemble} of supernovae
at a particular redshift.  
Implicitly, {\em individual} supernovae
are treated as either detectable or not, i.e., dust effects are
considered negligible or total;
our calculation treats total, catastrophic losses of
supernovae using this $f_{\rm dust}$ formalism.
We separately include the effect of partial extinction due to dust,
where the apparent magnitude of supernova is reduced 
but still visible, as discussed in the next section.
Of course in reality, all supernovae will experience some level of
extinction in their host galaxies, with the {\em distribution} of
host-galaxy extinctions changing with redshift.
A more detailed study of dust effects on supernovae (and uses of
supernovae to quantify and calibrate these effects) would
be of interest for further investigation;
see discussion in \S \ref{sect:dustfuture}.

\subsection{Supernova Observability at Cosmic Distances}

\subsubsection{The Supernova Luminosity Function}
\label{sect:snlf}

Locally observations of core-collapse supernovae 
reveal diverse light curves,
with a wide range in peak luminosity, and very different
evolution after maximum brightness.
The vast majority of supernovae discovered by
synoptic surveys will
lie at cosmological distances, and thus
will be detectable mostly near their maximum luminosity.
Thus we will focus on the observed distribution of
peak brightness, and the timescales on which
supernovae sustain it.

The distribution of peak absolute magnitude  $\Mpeak$ 
is given by the supernova luminosity function
$\snlf = \snlf(\Mpeak;z)$ which may have a redshift dependence;
we choose a normalization such that at any $z$,
$\int \snlf(M;z) \ dM = 1$, with this, we may write
the cosmic supernova rate per absolute peak magnitude as
\beq
\label{eq:csnlf}
\frac{dN_{\rm SN}}{dV_{\rm com} \, dt_{\rm em} \, d\Mpeak} \equiv  \csnr(z) \ \snlf(\Mpeak)
\eeq
where here and throughout the possible redshift dependence
of the luminosity function is understood.

\citet{richardson}
find the best-fit formulae for the supernova
peak luminosity functions in $B$-band for different types of 
supernovae based on their tabulation of 
279 supernovae of all types, for which absolute magnitudes were
available at peak brightness.
Of these, there were 168 events of all core-collapse types:
II-P,L,n and I-ab.
For each type, \citet{richardson} fit the observed $B$-band
absolute magnitude distributions with gaussian profiles,
in some cases including two profiles where the data
suggested ``bright'' and ``dim'' subclasses.
Their results 
provide the basis for the luminosity functions used in this paper.

Note that we use the {\em observed} distributions
rather than intrinsic, dust-corrected versions.
Thus we automatically include the mean extinctions
(ranging from $A \sim \mag{0.1}-\mag{0.3}$ for different types)
found at low redshift. 
The redding effect due to dust 
(ranging from $E \sim \mag{0.02}-\mag{0.26}$ across different
bands and redshifts)
is also added, 
based on the information given by \citet{kl}. 
As noted in the previous section,
catastrophic losses of supernovae due to large
extinction and its possible evolution at high redshift is treated
separately via our $f_{\rm dust}$ parameter.

We adjust the \citet{richardson}
distributions in two ways.  First, we converted from their Hubble constant of
$h = 0.6$ ot our adopted value $h=0.71$.
More importantly, we assume that each gaussian is a good
representation of the data around the peak, but we do
not allow the wings to extend arbitrarily far.  Instead,
we cut off the distributions at $|M - M_{\rm mean}| > 2.5 \sigma$, where
no data exist in the  \citet{richardson} sample.
We introduce these cutoffs in order to avoid extrapolating to
very rare, bright events which in a large survey could extend the
redshift reach considerably.  Below (\S \ref{sect:results}) we 
discuss the effect of this cutoff and its effect on the
predicted supernova redshift range.

\subsection{Supernova Discovery in Magnitude-Limited Surveys}
\label{sect:snmlim}

Surveys will discover supernovae 
monitor lightcurves in one or more passbands
Here we will adopt the SDSS $ugriz$ photometric
system, which uses AB magnitudes \citep{fukugita}.

The light curve of any supernova will suffer redshifting
and time dilation effects. 
For passband $x$ we put
\beq
\label{eq:distmod}
m_x-M_x = 5 \log \pfrac{d_L(z)}{d_0} + K_x(z) + A_x(z)
 \equiv \mu(z) + K_x(z)  - 1.086 \ln f_{\rm dust}
\eeq
with $d_L$ the luminosity distance and 
$\mu(z)$ is the usual distance modulus with $d_0 = 10$ pc.
The dust extinction $A$ is included via
the factor $f_{\rm dust}$ (eq.~\ref{eq:dust}).
The $K$-correction accounts for redshifting of the supernova
spectrum, and is discussed in Appendix \ref{sect:K-correction}.

As noted above, at each redshift
the effect of dust will be to obscure some fraction
of supernovae.  The remaining
unobscured events will have apparent 
$x$-filter magnitudes of $m_x=M_x+\mu(z)+K_x(z)$.
The expected $m_x$ {\em distribution} thus
reflects the underlying distribution of absolute magnitudes
$M_x$.
Since the \citet{richardson} supernova luminosity function we use 
is in the $B$-band, we need to find the corresponding 
$B$-band magnitude in order to find the 
right corresponding number of supernovae;
this transformation to $m_B$ is straightforward 
and is given by $m_x = m_B + \eta_{xB}$,
where
\beq
\eta_{xB} =
  - 2.5 \log\frac{\int^{x_f}_{x_i} F(\lambda)S_x(\lambda)d\lambda}
                 {\int^{B_f}_{B_i} F(\lambda)S_B(\lambda)d\lambda}
  +\rm zeropoint \; correction
\eeq
is a color index which translates between the
$x$ and $B$ magnitudes in the rest frame, and zeropoint correction is 
the correction for different zeropoint of the SDSS magnitude system 
and the Johnson magnitude system.
For the
spectral shapes $F(\lambda)$ we use the
prescriptions of \cite{dahlen} as described in 
Appendix \ref{sect:K-correction}.

The absolute $x$-band magnitude distribution of unobscured supernovae
at redshift $z$ is $\snlf[M_x-\eta_{xB}]$,
where $\snlf$ is the luminosity function 
in $B$-band as tabulated by \citet{richardson}.
Therefore the distribution of a certain type of supernova {\em apparent}
magnitudes $m_x$ in $x$-filter is 
$\snlf[m_x-\mu(z)-K_x(z)-\eta_{xB}]$,
Thus 
the fraction of all (unobscured) supernovae at $z$ 
which fall within the survey  $x$-band magnitude limit $\mlim$
is a sum over the luminosity functions for all core-collapse types:
\begin{align*}
f_{\rm maglim} (z) &= \sum_{\rm types} \frac{\int^{\mlim{x}} 
  \snlf[m_x-\mu(z)-K_x(z)-\eta_{xB}] \ dm}{\int \snlf(m) \ dm} \\
  &= \sum_{\rm types} \frac{\int^{\mlim-\mu(z)-K_x(z)-\eta_{xB}} \snlf(m^\prime) \ dm^\prime}{\int \snlf(m) \ dm} \\ 
  &\equiv f_{\rm snlf}[<M_{\rm lim}(z,\mlim)]
\end{align*}
which is the cumulative fraction of supernovae whose absolute magnitude is 
brighter than 
\beq
M_{\rm lim}(z,\mlim) = \mlim-\mu(z)-K_x(z)-\eta_{xB}
\label{eq:Mlim}
\eeq

To develop some intuition, suppose the
supernova peak brightnesses lie in a range
$\Mpeak \in (M_{\rm bright},M_{\rm dim})$, and ignore for
now the effects of dust.
Then for low redshifts such that
the absolute magnitude limit $M_{\rm lim}$ from eq.~(\ref{eq:Mlim})
is fainter than $M_{\rm dim}$, 
we can expect to see {\em all} supernovae, and $f_{\rm maglim} = 1$.
For these redshifts, we can study the entire supernova luminosity function
and test whether it varies with redshift.
On the opposite extreme, for high redshifts such that
$M_{\rm lim}$ is dimmer than $M_{\rm bright}$ 
we can see {\em no} supernovae, so 
$f_{\rm maglim} = 0$;
this then defines the survey redshift cutoff (for fixed $\mlim$).
Finally, for intermediate $z$ such that
$\mlim - M_{\rm dim} < \mu(z)+K_x(z)+\eta_{xB} < \mlim - M_{\rm bright}$,
we have $0 < f_{\rm maglim} < 1$;
over these redshifts the survey samples the bright end of the
supernova luminosity function. 

Both magnitude limit and dust extinction reduce
the expected supernova detection, and do so
independently of each other.  Consequently, we can
find the net supernova detection probability
by simply taking the product of the individual
factors:
\beq
f_{\rm detect}(z; \mlim) = f_{\rm maglim}(z;\mlim) \ f_{\rm dust}(z)
\eeq

\begin{figure}[!h]
\includegraphics[width=0.4\textwidth,angle=90]{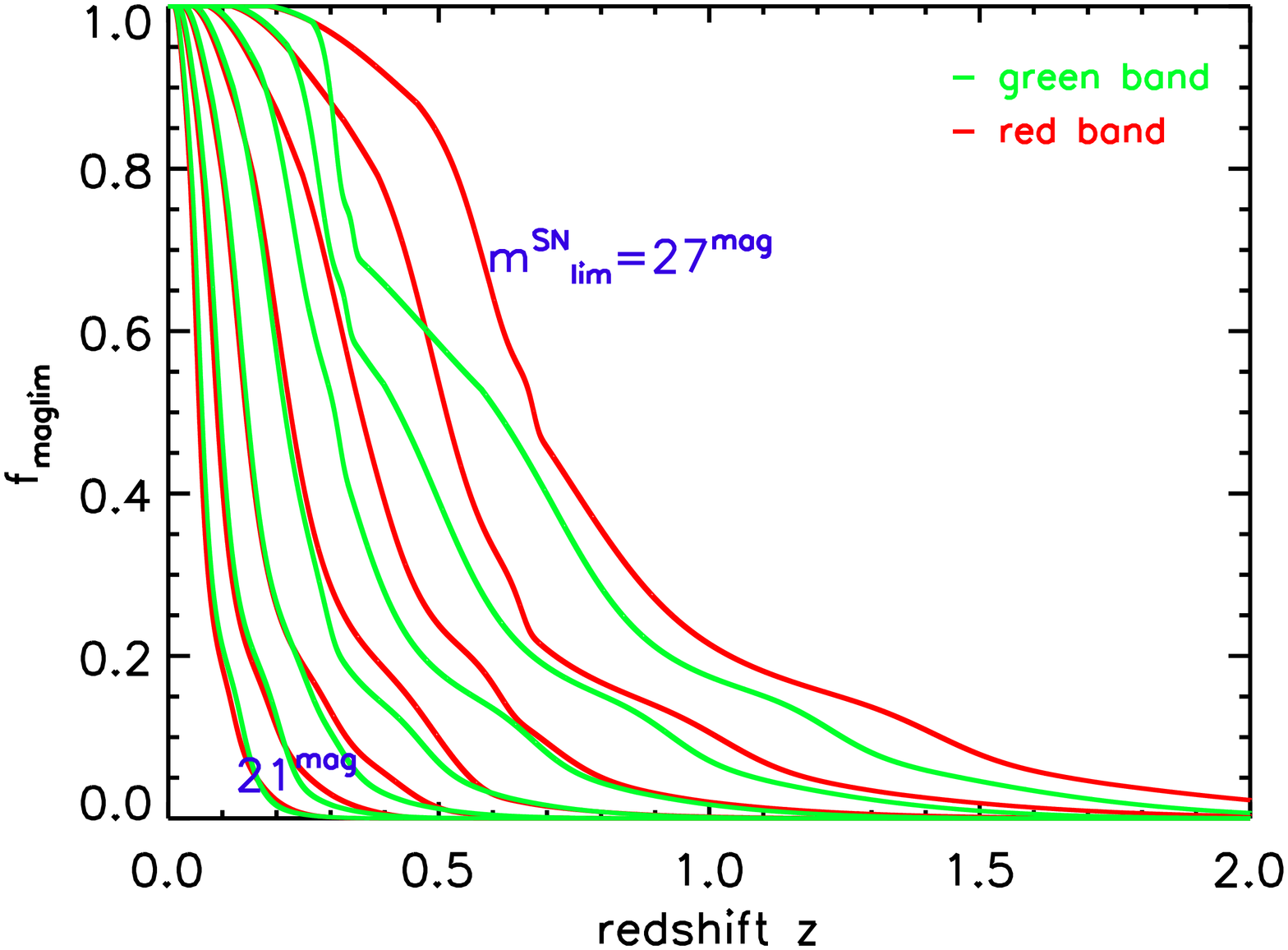} 
\includegraphics[width=0.4\textwidth,angle=90]{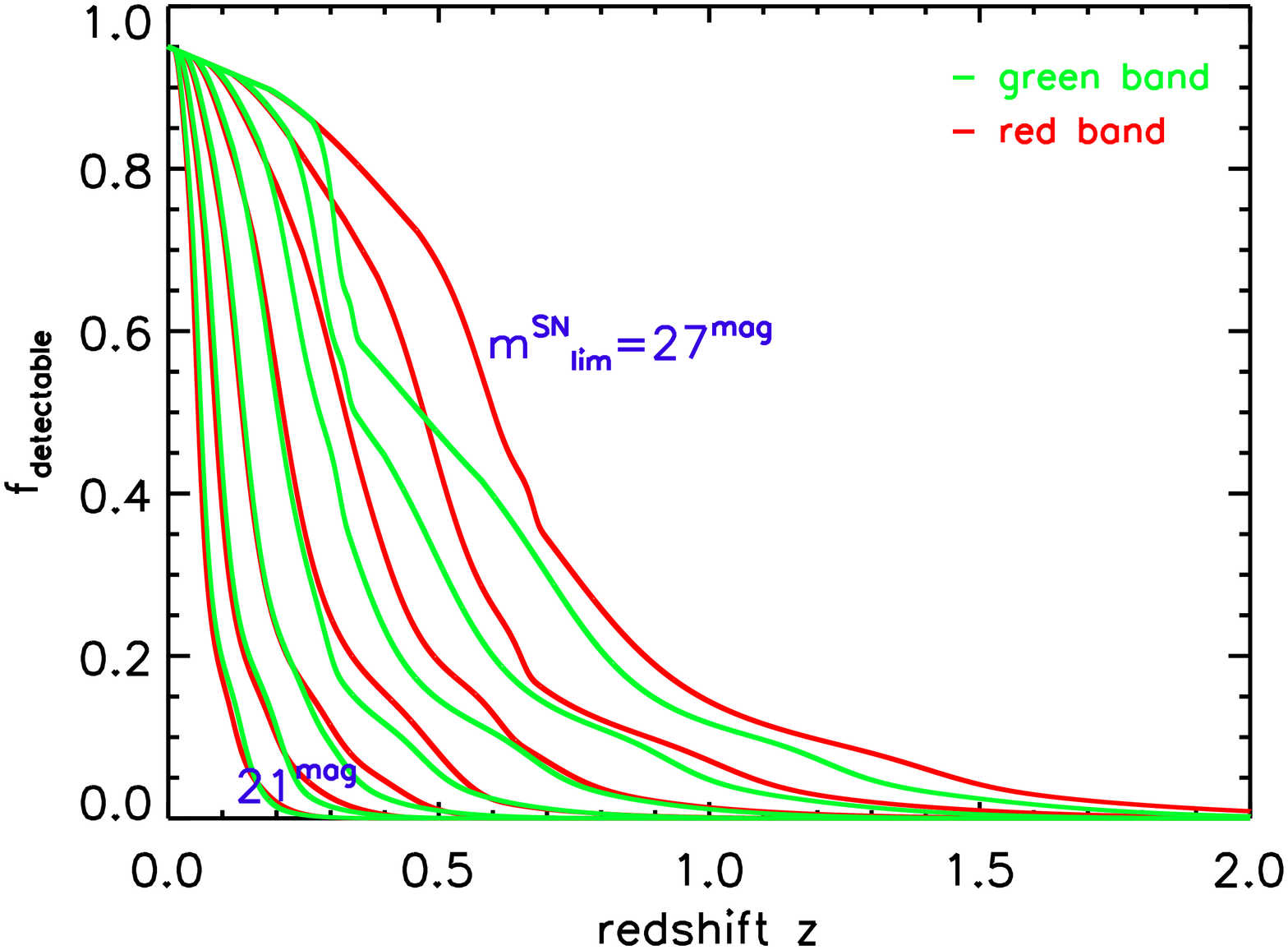}
\caption{(a)
The fraction of supernovae detected based on different survey
deepness in $g$ and $r$ bands, with $\mlim$ ranging from $\mag{21}$ down to $\mag{27}$;
effects of dust obscuration are not included.
(b) As in (a), but including the effects of dust obscuration
strongly evolving with redshift as modeled by eq.~\pref{eq:dust}.
\label{fig:fdetectable}
}
\end{figure}

Figure \ref{fig:fdetectable}
shows the resulting detectable fraction of supernovae.
The left panel  
shows the shape of $f_{\rm maglim}$ for the $g$
and $r$ bands.
At redshifts close to zero, $f_{\rm maglim} \approx 1$
which means that almost all supernovae are detected in the local
universe. And it approaching to zero at high redshift, which reflects
the fact that no supernovae can be detected at high redshift because
of the survey deepness. Note that $g$ and $r$ bands are competitive
for $\mlim \leq 24$, but for higher $\mlim$, $f_{\rm maglim}$ in $g$-band drops a lot faster
than those in $r$-band especially around z $\sim$ 0.4,
which is cause by the effect of $K$-correction.
The figure also shows that for higher $\mlim$,
$f_{\rm maglim}$ decays less rapidly. 
The right panel 
shows $f_{\rm detect}$
for different $\mlim$, using
our adopted dust model (eq.~\ref{eq:dust}). 
We see $f_{\rm detect}$ shows the same trend as  $f_{\rm maglim}$ 
except the detectable fraction is reduced due to dust 
and we can no longer observe all supernovae even 
in the local universe.
It is also clear to see that going to fainter $\mlim$ 
significantly boosts the detectable fraction at
high redshift.
For $\mlim=\mag{23}$, $f_{\rm detectable}$
is almost zero at redshift $z \sim 1$ for both $g$ and $r$ bands.
But going to $\mlim=\mag{26}$, 
$\sim55\%$ of the supernovae  at redshift $z \sim 1$
remain visible both the $g$ and $r$ bands.

This means that deeper surveys (and/or scanning modes
in which smaller areas are scanned more deeply) will 
probe supernovae out to much  higher redshifts.
Deep survey modes will also probe a much wider regime of the
supernova luminosity function and light curves over a broad range of
cosmic epochs, thus testing for redshift evolution in
supernova properties.
The clear lesson is that the scan $\mlim$ is critical
in determining the quality and reach of
the supernova science.  In particular, 
we urge that scans strategies include modes which push
$>\mag{1}$ deeper than the all-sky depth.

\subsubsection{Supernova Light Curves}

The observed population of 
core-collapse supernovae shows a broad range of timescales and
time histories in their decline from peak brightness
\citep[e.g.,][]{db,ls}.
The amplitude and time behavior of these curves
encodes a wealth of information about the
underlying physics of the supernovae as well as
their interaction with the circumstellar and interstellar
medium.

Empirically, light curves broadly fall into phenomenological
categories, those whose magnitudes decline in a relatively
steep, linear way (Type II-L) and those which
linger near peak brightness with a relative plateau
in magnitude (Type II-P).
\citet{patat93} compiled 51 Type II light curves,
and analysis in 
\citet{patat94}
showed that plateau-type supernovae typically decline from peak
brightness at rates which vary
the range $(\mag{0.7} - \mag{3.1})/100$ days,
while linear-type events typically have
$(\mag{3.9}-\mag{5.7})/100$ days.
Unfortunately, the lightcurves available at the time of
these studies were poorly sampled near the peak itself,
where the behavior is most critical for our purposes.

Fortunately, subsequent data, particularly
using {\it Swift}, gives a clearer picture of the
early light curves
for a few events.
For plateau event SN 2005cs,
data in \citet{pastorello}
show that $\sim 15$ days after peak brightness,
the supernova dimming was strongly depending on passband:
$\Delta M_{15}(U) \simeq \mag{1.8}$,
$\Delta M_{15}(B) \simeq \mag{0.7}$,
$\Delta M_{15}(V) \simeq \mag{0.18}$,
and
$\Delta M_{15}(R) \sim \mag{0.1}$.
Another Type II-P event, SN 2006bp,
after $\sim 13$ days declined by $\sim \mag{1}$ in $U$,
but within errors was essentially constant in $B$ and $V$
\citep{dessart}.
For Type Ib, the recent event SN 2008D was seen from 
shock breakout \citep{modjaz};  
after dropping from this brief initial outburst, the
flux increased for $\sim 15$ days to a maximum.
Afterwards, the brightness decline rates lengthen with wavelength,
with a drop of $\Delta M \sim \mag{1}$ 
after $\sim 10$ days in $U$-band, but
after about 15 and 20 days in $B$ and $V$ respectively.

These multicolor data show that brightness decline in
$V$ and longer passbands comparable to if not slower than
the typical range of $\Delta M_{15} \sim \mag{1} - \mag{2}$
found in Type Ia events
\citep{phillips}.  This implies that surveys
timed for Type Ia discovery will automatically be
well-suited and possibly even better-sampled for core-collapse
events.
In particular, we will find below that the $r$ and also $g$ passbands
are the most promising for survey supernova detections.
Thus, if cosmic supernovae follow the behavior
of these local events, we expect that the light
curves will remain within, e.g., $\Delta m \simeq \mag{0.5}$ of peak
brightness (a factor 1.5 in flux) for a timescale 
of at least a week.  In some cases this timescale will
be longer, and possibly
also with detections in the rising phase.

For synoptic surveys to detect core-collapse
supernovae near their peak brightness, 
the cadence needs to be shorter than
the (observer-frame) brightness decline time.
Thus weekly revisits are sufficient for marginal detections,
and cadences of $\sim 3-4$ days will often see the event 
three or more times.
In the cases of plateau events,
the supernova should remain near peak brightness for
many such revisit times.
Furthermore, due to cosmological time dilation effects,
the observed brightness decline timescale
$\tau_{\rm obs} = (1+z) \tau_{\rm rest}$
is increased by a factor of $1+z$,
which extends the detection window
and offers a greater opportunity to recover
a well-sampled lightcurve.
Also, we see that color evolution 
is not strong in $V$ and $R$ bands.
The UV and blue do fade more rapidly,  and the supernova reddening
depends on the type.
For events where bluer rest-frame colors are available,
this might be a useful means of photometrically determining supernova type.

\section{The Cosmic Core-Collapse Supernova Rate:  
Forecasts for Synoptic Surveys} 
\label{sect:forecasts}

In this section  we will work out general formalism
for supernova observations by synoptic surveys.
We then apply this formalism to specific current and proposed surveys

\subsection{Connecting Cosmic Supernovae and Survey Observables}

\subsubsection{General Formalism}

It is useful to define a differential supernova detection rate
per unit redshift, solid angle, and apparent magnitude in $x$-band:
\beq
\label{eq:dNdtdzdOdm}
\frac{dN_{\rm SN,obs,x}}{dt_{\rm obs} \, dz \,  d\Omega \, dm} 
 = \csnr(z)
 \ \frac{r(z)^2}{1+z} \frac{dr}{dz} \ f_{\rm dust}(z) \ \snlf[m_x-\mu(z)-K_x(z)-\eta_{xB}] 
\eeq
This expression
adds the effects of supernova luminosity 
(cf eq.~\ref{eq:csnlf})
and of dust obscuration
(eq.~\ref{eq:dust})
to the ideal rate of eq.~\pref{eq:idealrate}.
Throughout, we will for simplicity refer to the entire core-collapse
supernova population, but the formalism could equally well
distinguish the various core-collapse types,
and compute the rates of each.
An example of such a treatment is the 
\citet{scannapieco} study of the rate and detectability of
pair-instability supernovae.

The differential rate in eq.~\pref{eq:dNdtdzdOdm}
relates the observables in a synoptic survey 
to underlying properties of cosmic supernovae.
As such, a wealth of information can be recovered
by a good statistical sample of supernovae
over a redshift range:
one probe different terms and their underlying physics.
For example, at fixed $z$, the range of observed
supernova magnitudes in $x$-band $m_x$ probes the supernova
peak luminosity function
$\snlf(M;z)$ at magnitudes $M_x=m_x-\mu(z)-K_x(z)-\eta_{xB}$.
Comparing these results at different redshifts
with local determinations
can reveal any redshift- and/or environment-dependence 
in the core-collapse supernova luminosity function.

Another aspect of cosmic supernovae probed by
synoptic surveys, and central focus of this paper,
is the cosmic supernova rate.
Whereas the supernova luminosity function can be determined
from the distribution of supernova magnitudes at the {\em same} redshift,
the cosmic supernova rate comes from the distribution of supernova
{\em counts} across {\em different} redshifts.
The observed differential rate for supernovae 
of all magnitudes in the $x$-band is 
\beq
\Gamma_{\rm SN,obs,x}(z) 
  \equiv  \frac{dN_{\rm SN,obs,x}}{dt_{\rm obs} \, dz \,  d\Omega} 
 =   \int^{\mlim} dm \ \frac{dN_{\rm SN,obs,x}}{dt_{\rm obs} \, dz \,  d\Omega \, dm}\\
  =  \csnr(z) \ f_{\rm detect,x}(z;\mlim) \ \frac{r(z)^2}{1+z} \ \frac{dr}{dz}
\eeq
Note that this is the idealized rate of eq.~\pref{eq:idealrate}
reduced by the detection in $x$-band $f_{\rm detect,x}$.

\begin{figure}[!h]
\includegraphics[width=0.9\textwidth]{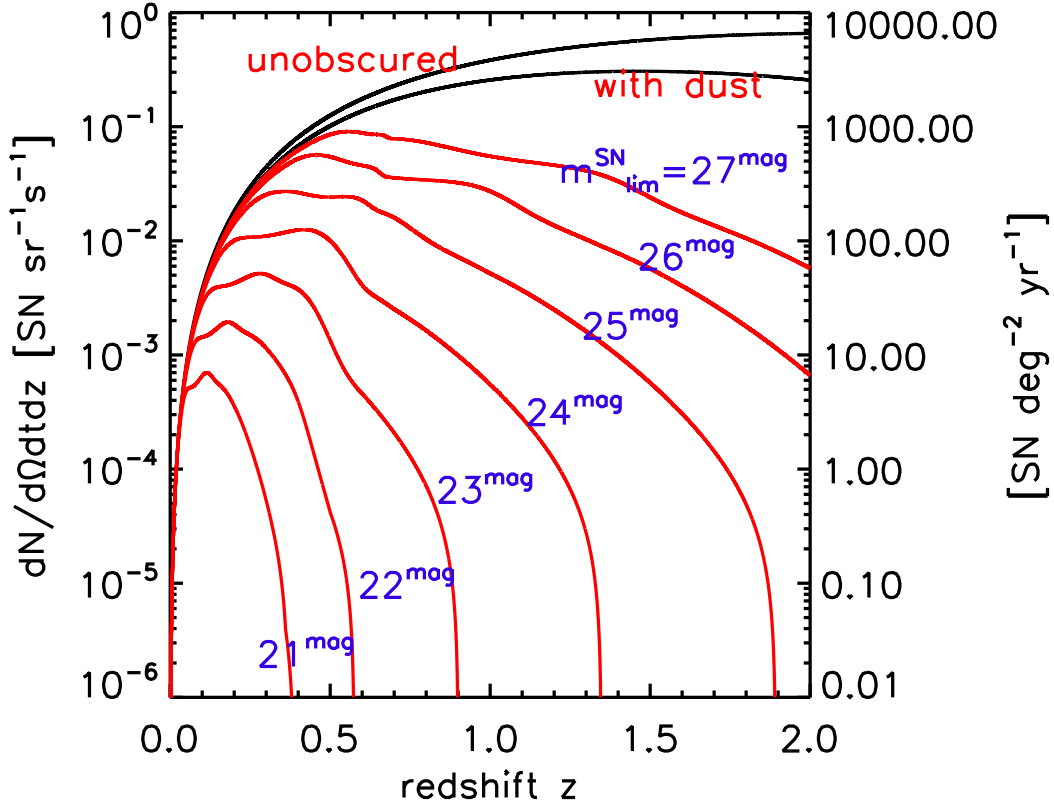}
\caption{The cosmic supernova detection rate in $r$-band,
expressed in number of events per solid angle per time,
shown a function of redshift.
The curve labeled ``unobscured'' ignores both effects of
dust extinction
or the flux limit of the survey (i.e., $f_{\rm detect} = 1$).
The curve labeled ``with dust'' includes dust extinction only,
but with $\mlim = \infty$.
The remaining curves
are for surveys with $\mlim$ as labeled,
and include dust extinction.
Note that the vertical axis is shown
both in units of events per second per
steradian (left scale) and events per year per
square degrees (right scale).
}
\label{fig:dNdodtdz_r}
\end{figure}
 
One can get a sense of the orders of magnitude in play
via the definition of
a dimensionful scale factor
\beqar
\label{eq:scale}
\Gamma_{\rm SN,0} & = & \csnr(0) \ d_H^3
  =7.2 \times 10^{6} \ {\rm events \ yr^{-1} \ sr^{-1}} 
 \pfrac{\csnr(0)}{10^{-4} \ {\rm yr^{-1} \, Mpc^{-3}}}   \\
 & = & 0.22 \ {\rm events \ sec^{-1} \ sr^{-1}}  \pfrac{\csnr(0)}{10^{-4} \ {\rm yr^{-1} \, Mpc^{-3}}}   \\
 & = & 2.2 \times 10^{3} \ {\rm events \ yr^{-1} \ deg^{-2}}
  \pfrac{\csnr(0)}{10^{-4} \ {\rm yr^{-1} \, Mpc^{-3}}}   
\eeqar
We may then define 
a dimensionless distance $u(z) = r(z)/d_H$, 
with $d_H = c/H_0$ the Hubble length,
and write
\beq
\Gamma_{\rm SN,obs}(z) = 
  \Gamma_{\rm SN,0} 
  \ \frac{\csnr(z)}{\csnr(0)} 
  \ \frac{u(z)^2}{1+z} \frac{du}{dz} 
  \ f_{\rm detect,x}(z,\mlim) 
\eeq

Figure \ref{fig:dNdodtdz_r} plots the
observed supernova rate $\Gamma_{\rm SN,obs}$ per
solid angle in $r$-band.
For comparison, we show the idealized cases of $\mlim = \infty$
and $f_{\rm dust} = 0$,
as well as realistic cases in the presence of dust
and with different $\mlim$.
The {\em amplitudes} of the curves in
Figure \ref{fig:dNdodtdz_r} confirm the large numbers of events 
expected from eq.~\pref{eq:scale}.

The {\em shapes} of the curves
can also be readily understood.
At low redshifts, the surveys see most
of the supernovae that occur--i.e., the entire luminosity
function is sampled; cf Figure \ref{fig:fdetectable}.
Hence at small $z$, the supernova sample is
simply limited by the cosmic volume within $z$:
$\Gamma \propto dV_{\rm com}/dz \sim r_{\rm com}^2 \, dr_{\rm com}/dz \sim z^2$
Thus the detection rate initially 
rises quadratically with $z$; this volume effect is essentially
independent of survey magnitude limit, as we see by the
overlap of the curves in this regime.

In the high redshift limit, several effects
act to suppress supernova detectability.
At $z>1$, $r_{\rm com}$ rapidly saturates at
the comoving horizon scale, and nearly all 
observable cosmic volume is sampled; in this regime,
the volume factor {\em decreases}
as $dV_{\rm com}/dz \sim dr_{\rm com}/dz \sim 1/H(z) \sim (1+z)^{-3/2}$.  
In addition, time dilation effects become large and add another factor
of $(1+z)^{-1}$.
For these reasons, even the idealized (unobscured, $\mlim=\infty$)
rate drops.
Moreover, in some models \citep[such as that of][]{cole},
the CSNR itself is intrinsically expected to drop
after a peak, perhaps somewhere in the range $z \sim 1-3$.
On top of this, the effects of dust obscuration
become large at $z \ga 1$ and removes further
supernovae in this range.  Finally, a finite survey magnitude limit
truncates still more events at high $z$.

The combination of the low-redshift rise and high-redshift
drop acts to create a peak in supernova detectability.
The position of the peak is sensitive to the CSNR itself,
and the details of dust
obscuration.
But the peak position and amplitude
are also both very sensitive to
the survey magnitude limit;
both rise sharply 
as survey depth $\mlim$ increases. 
This illustrates a key conclusion which will be
manifest in several other ways below:
{\em for discovery of core-collapse supernovae
at high redshifts, 
the most important aspect of a synoptic survey is its
limiting magnitude; investment in deep scan
modes ($\mlim > 24$ mag) will
reap substantial rewards.}

\begin{figure}[!h]
\includegraphics[width=0.9\textwidth]{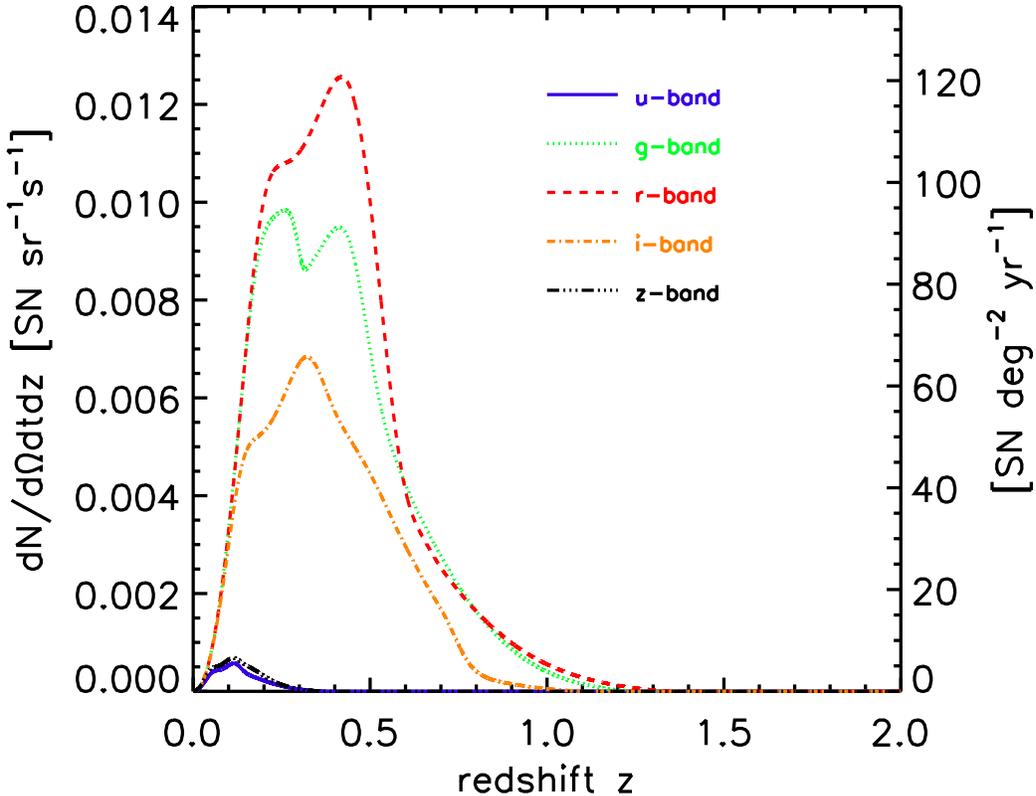}
\caption{Number of supernovae per year per solid angle per redshift 
with $\mlim = \mag{24}$ in different bands.}
\label{fig:dNdodtdz_band}
\end{figure}

Figure~\ref{fig:dNdodtdz_band} shows
the same supernova rate redshift distribution
as in Fig.~\ref{fig:dNdodtdz_r}, but for
the five $ugriz$ passbands with SDSS filters and efficiencies.
For each band we fix $\mlim = \mag{24}$.
We see that the discovery rate is the highest in $r$ for
essentially all redshifts, with $g$-band counts very nearly
the same except around the peak at $0.2 \la z \la 0.6$. 
The relative smallness of the counts in other
bands traces back predominantly to low
detector efficiency in $i$ and $z$,
and redshifting effects for $u$.
The upshot is that for synoptic surveys,
$r$ and $g$ bands are (in that order) clearly the most
promising for supernova search.

\begin{figure}[!htb]
\includegraphics[width=0.5\textwidth]{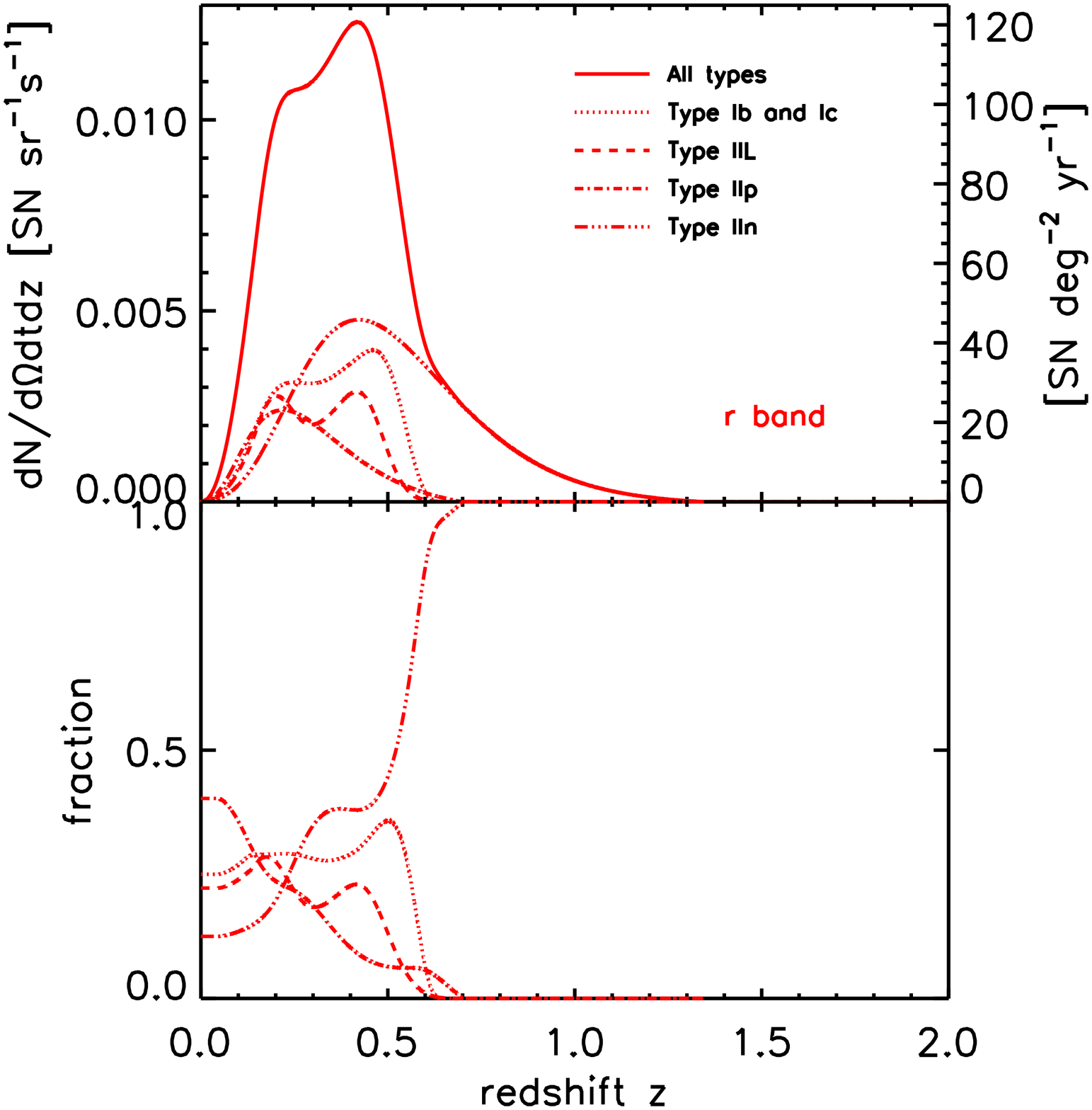}
\includegraphics[width=0.5\textwidth]{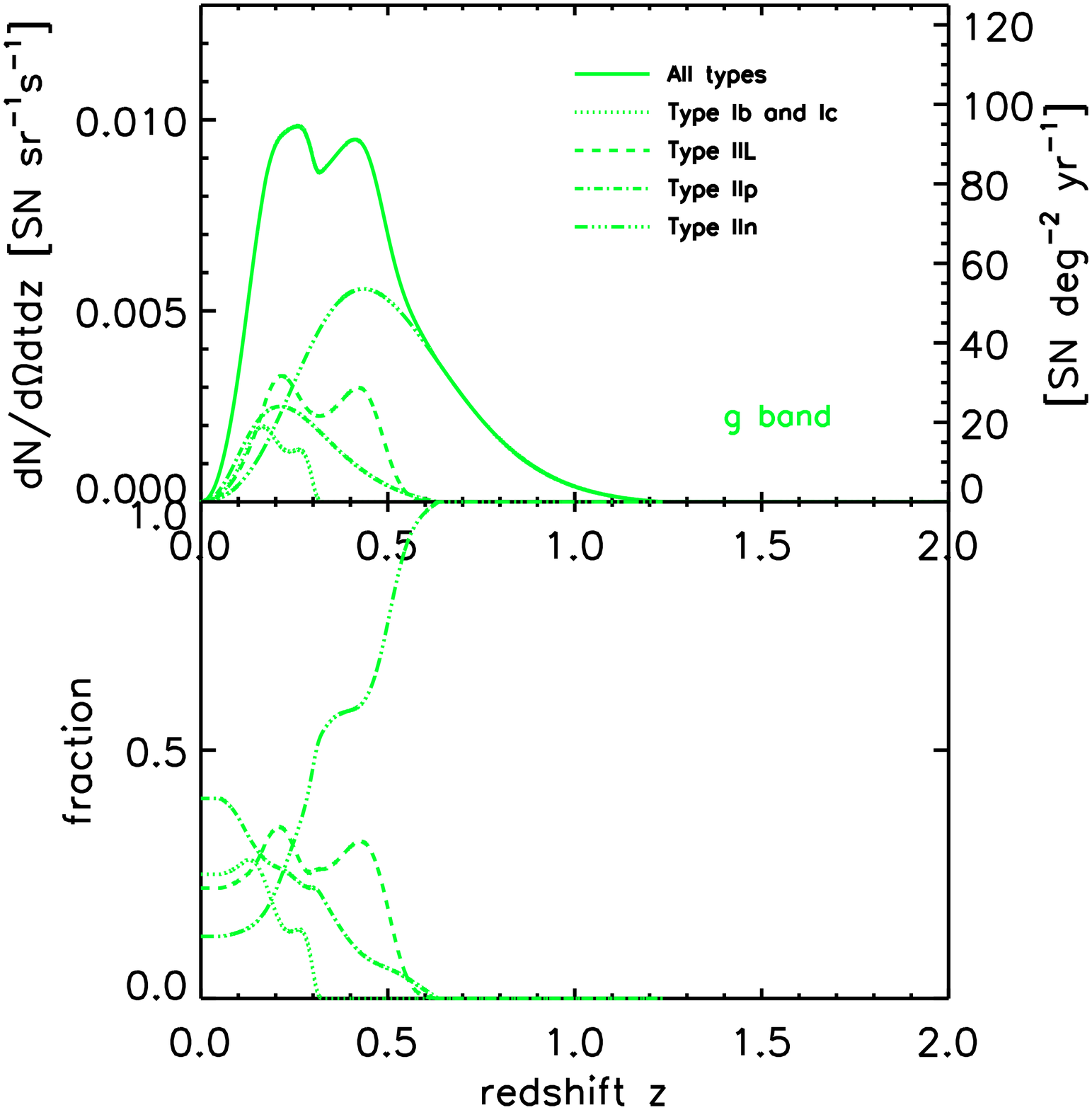}
\caption{Supernova rate redshift distribution, as in 
Fig.~\protect{\ref{fig:dNdodtdz_band}},
broken down by core-collapse type.  
Results shown for (a) $r$ band, and
(b) $g$ band;
both have $\mlim = \mag{24}$.
{\em Top panels:} 
detection rate distribution per subtype;
{\em bottom panels:} 
fraction of each subtype rate relative to the total.
We see that intrinsically bright Type IIn events dominate 
the counts at high redshift ($z \gtrsim 0.5$) and thus determine the redshift 
reach for core-collapse discovery.
\label{fig:dNdodtdz_types}
}
\end{figure}

We have thus far shown the {\em total}
supernova rate redshift distribution, summed over all core-collapse subtypes.
Figure~\ref{fig:dNdodtdz_types}
illustrates how the different subtypes contribute to the aggregate.
Here we fix $\mlim = \mag{24}$ and show results for the $r$
and $g$ bands.
It is worth recalling that we have assumed the
low-redshift \citet{richardson} determination of
luminosity functions and type distributions holds for
all redshifts.
In this scenario, we see that in both bands,
Type IIn events
give the largest contribution to the signal at $z \ga 0.3$, and 
totally dominate the counts  at $z \ga 0.6$.
This is expected, since
it is the intrinsically brightest core-collapse subtype.
Thus the redshift reach of supernova discovery (and
associated results such as the CSNR) in synoptic surveys
will depend sensitively on nature Type IIn events at 
$z \ga 0.6$.
It will thus be crucial to determine whether these events show evolution
in their luminosity function and/or relative fraction
of core-collapse events with redshift (e.g., via metallicity effects).
Also, it is worth noting that the \citet{richardson}
luminosity function we have used is relatively narrow.
As noted recently by \citet{cooke},
some Type IIn events have now been observed with luminosities
far above the range of values we consider. If so,
then the redshift range of synoptic surveys could thus extend
significantly further than in our estimates.

Figure~\ref{fig:dNdodtdz_types} further
predicts that the other core-collapse types
should have observably distinct redshift ranges
in different bands,
again assuming no evolution in luminosity function or type
distribution.
The upper panels of Fig.~\ref{fig:dNdodtdz_types}
show the individual subtype detection rates,
as well as their sum.
Type II-L events
have simlilar behavior
in both $r$ and $g$ bands, 
peaking at $z \sim 0.45$ then rapidly dropping off.
Although Type II-P events are the largest core-collapse subtype
in the \citet{richardson} sample,
they are also by far 
the intrinsically dimmest,
$\sim \mag{1}-\mag{3}$ fainter than the other types.
We thus find that Type II-P have a smaller redshift range than 
Type II-L and IIn events.
The counts and redshift range of Type Ib and Ic events
are notably different in the two passbands.
This traces to the effects of UV lineblanketing which
removes blue flux; thus at high redshift
the $K$-correction first shifts photons out of the $g$
band, with the $r$-band signal surviving until
higher redshift.
Note also that the ``bright'' and ``normal'' Type Ib and Ic
events lead to the double-peaked structure in their redshift
distribution.  

The lower panels of Fig.~\ref{fig:dNdodtdz_types}
shows our forecast for subtype fraction detected
as a function of redshift, i.e., the ratio of each subtype
rate to the total.
At $z=0$, the subtype fractions go to the
observed local values we have adopted from \citet{richardson},
as required by our model design.
For $z \sim 0-0.2$, we see that all subtypes make
significant contributions tot the total, and thus for
this redshift range, the
sharp rise in the total detection rate (top panel)
is due to contributions from all subtypes.
The features around the maximum in the total rate
($z \sim 0.2-0.5$) are due to
the interplay between the rise of the Type IIn events
and the successive dropout of the other types.
Finally, we see that for $z \sim 0.5$, Type IIn events 
essentially completely set the total rate.

Because the highest-redshift detections 
will be dominated by Type IIn events, the nature of and evolution of
this subtype will play a crucial role in setting the 
high-redshift impact 
of surveys for core-collapse events,
as also pointed out by \citet{cooke}.
As we have noted, intrinsic evolution of the Type IIn fraction
of core-collapse events would directly change--and be written into--the 
high-redshift signal.  But at present, the uncertainties are 
very large even when evolution issues are set aside.
Namely, published data are as yet very uncertain concerning
the local, $z \approx 0$ fraction of core-collapse events
which explode as Type IIn.  Our forecasts use the
\citet{richardson} sample which finds 
9 Type IIn events out of 72 core-collapse events,
for a fraction of 12.5\%.
However, this {\em discovery} fraction
are very uncertain.  For example, 
the prior work of \citet{dahlen} 
compiled their own core-collapse discovery statistics,
and adopted a Type IIn event fraction of $2\%$,
while noting that \citet{cappellaro} recommend
a Type IIn fraction of $\sim 2-5\%$.
Because the high-redshift core-collapse detections
will be dominated by Type IIn events, if these
values better reflect the intrinsic fraction, this
would dramatically reduce our predicted detection
rates for $z \ga 0.5$ by factors of $\sim 2-6$,
and thus also reduce the maximum redshift at which core-collapse
events can be seen in surveys.
Clearly, the small numbers available when all of these
compilations were made
render the Type IIn fraction estimates uncertain;
indeed, to a lesser extent the estimates for the more
common core-collapse types suffer similar problems.

In light of the uncertainties in the 
\citet{richardson} and prior compilations, 
it is worth noting that considerably more
supernova data already exists.  A detailed, systematic study of the
luminosity function and intrinsic subtype fractions of
local events would be of the utmost value for
forecasts of the sort we have presented.  Moreover,
precise and accurate local measurements will play
an essential role
as a basis of comparison for the future medium- to high-redshift
data, in order to empirically probe for evolution within
and among the core-collapse subtypes.

\subsubsection{Unveiling the Cosmic Core-Collapse Supernova Rates}

As noted above, synoptic surveys will revolutionize our
understanding of the CSNR because they will directly
determine the rate through {\em counting}.
We now are in a position to determine the supernova
counts for realistic (magnitude-limited, dust-obscured)
surveys.  Using these, we can demonstrate how the
CSNR can be extracted. We can further determine its statistical
uncertainty and the impact of survey depth and sky coverage.

Consider a survey with scan area $\oscan$ and limiting magnitude $\mlim$,
the total number of supernovae seen in $x$-band in time $\Delta t_{\rm obs}$, 
in a small redshift bins of width $\Delta z = z_f - z_i  \ll 1$
centered around $z = (z_f+z_i)/2$
is
\beqar
\Delta N_{\rm SN,obs,x} 
  & = & \oscan \Delta t_{\rm obs} \ \Delta z \ \Gamma_{\rm SN,obs,x}(z) \\
  & = & \oscan \Delta t_{\rm obs} \ \Delta z \ 
   \Gamma_{\rm SN,0,x}  \
   \frac{\csnr(z)}{\csnr(0)} \ \frac{u(z)^2}{1+z} 
  \frac{du}{dz} \ f_{\rm detect,x}(z;\mlim)
\label{eq:snnumber}
\eeqar
Thus we see that the cosmic supernova rate is directly encoded in our 
binned data. This means we can use the binned data to extract the supernova
rate:
\beq
\label{eq:surveycsnr}
\csnr(z) = \frac{1}{\oscan}  \ 
  \frac{1}{\Delta z} \ \frac{1+z}{u(z)^{2}} \frac{dz}{du} 
  \ f_{\rm detect,x}(z;\mlim)^{-1} 
  \ \frac{\Delta N_{\rm SN,obs,x}}{d_H^3 \, \Delta t_{\rm obs,x}}
\eeq
this result is a major goal of this paper.
Physically, we see that as we accumulate supernovae, i.e., as 
$\Delta N_{\rm SN,obs}$ fills out the redshift range accessible to the survey,
we obtain an ever better measure of the SN rate.

We can also compute the statistical uncertainty
in the CSNR 
derived from
counts in surveys.
The statistical error arises from the 
counting statistics in the supernova number.
Expressing this as a fractional error, we
have
\beq
\label{eq:csnr-sig}
\frac{\sigma(\csnr)}{\csnr} 
 = \frac{\sigma(\Delta N_{\rm SN,obs,x})}{\Delta N_{\rm SN,obs,x}}
 \approx  \frac{1}{\sqrt{\Delta N_{\rm SN,obs,x}}}
\eeq
But from eq.~\pref{eq:snnumber},
we see that $\Delta N_{\rm SN,obs,x}$ scales linearly
with the product of detected fraction and survey sky coverage,
as well as monitoring time and redshift bin width.
Thus we find the CSNR statistical error should scale as
\beqar
\label{eq:fracerror}
\frac{\sigma(\csnr)}{\csnr} 
  & = &    \frac{1}{\sqrt{\oscan
  \Delta t_{\rm obs} \ \Delta z \
   \Gamma_{\rm SN,obs,x}(z)}}
  \\
 & \propto & \frac{1}{\sqrt{f_{\rm detect,x}(z;\mlim) \, 
       \Delta t_{\rm obs} \, \oscan}}
\eeqar
Consequently, for a fixed redshift bin size $\Delta z$, 
the CSNR accuracy grows with
the product $\Delta t_{\rm obs} \, \oscan$, and implicitly with $\mlim$
via the detection fraction.
Thus survey sky coverage and magnitude limit (i.e., collecting area)
enter together, and we see the payoff of a large survey \'{e}tendue.

\begin{figure}[!hb]
\includegraphics[width=0.9\textwidth]{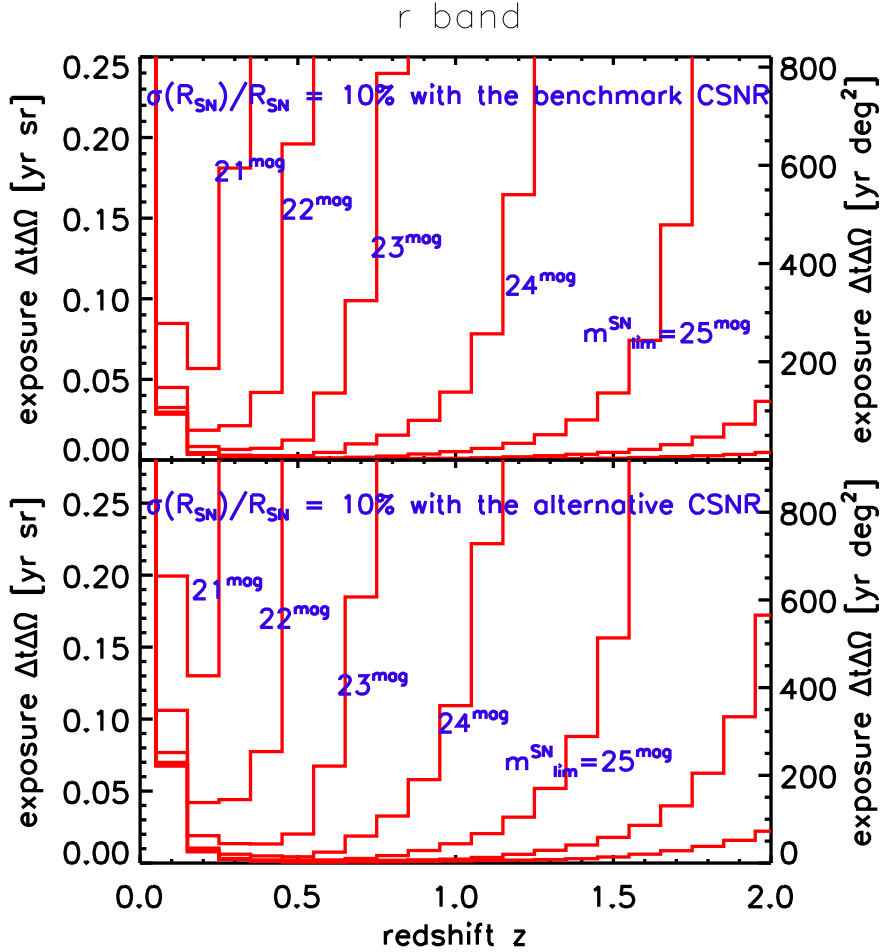}
\caption{Survey CSNR discovery parameter, i.e., the product of 
survey monitoring time and 
sky coverage $\Delta t \, \Delta \Omega$ needed to measure
the CSNR to a specified precision.  Data are binned in
redshift units of $\Delta z = 0.1$ vs. redshift.
{\em Top panel:} discovery parameter
needed to reach 10 $\%$ precision with our benchmark CSNR.
{\em Bottom panel:} discovery parameter
needed to reach 10 $\%$ precision with 
the alternative CSNR seen in Fig.~\ref{fig:csnr}.}
\label{fig:dtdomega_r}
\end{figure}

Thus, we can find the survey properties.
needed to achieve any desired precision 
in the CSNR at some redshift $z$.
For a fixed $\mlim$ and thus $f_{\rm detect}$,
monitoring time and sky coverage enter together
as the product
$\Delta t_{\rm obs} \ \oscan$.
Figure \ref{fig:dtdomega_r} shows 
the needed monitoring time $\Delta t_{\rm obs} \ \oscan$ 
to measure the CSNR to a statistical
precisions of $\sigma_{\rm stat}(\csnr)/\csnr = 10\%$,
and with different survey $\mlim$ in $r$-band.
In both panels we choose $\Delta z=0.1$ for the redshift bin size.
The two panels show our baseline and alternative 
CSFR. 
From these figures we can see that these two different
adopted CSNR behaviors both yield very similar
results for the survey CSNR detectability.

Again the shapes of the curves can be understood.
As shown in eq.~\pref{eq:fracerror},
that the precision at each bin scales inversely
with the supernova differential redshift distribution
as $\Gamma_{\rm obs}(z)^{-1/2}$.
Not surprisingly therefore,
the least monitoring is needed to measure the CSNR
for $z$ near the peak in the redshift distribution
On the other hand,
redshifts in the high- and low-redshift tails of
$\Gamma_{\rm SN,obs}$ require
increasing monitoring, eventually to the point
of unfeasibility.

Figure 
\ref{fig:dtdomega_r} makes clear that increasing 
$\mlim$ brings a huge payoff reducing the needed 
monitoring $\Delta t \ \oscan$. 
To achieve a 
$\sigma(\csnr)/\csnr < 10 \%$ 
precision at redshift $z=1$,
the monitoring becomes about 1000 times smaller in $r$-band 
if we 
increase $\mlim$ from $\mag{23}$ to $\mag{26}$. 
Clearly, for any survey, increasing $\mlim$ 
will drastically shorten the observing time 
needed for the high redshift supernovae.
In practice, given fixed survey lifetimes,
this means that $\mlim$ sets the maximum redshift reach
over which the survey may determine the CSNR
(via eq.~\ref{eq:Mlim}).

\subsection{Forecasts for Synoptic Surveys}
\label{sect:results}

\begin{table}
\caption{\label{tab:sn} Survey Discovery Potential
for Core-Collapse Supernovae
in $r$-band}
\begin{tabular}{c|cc}
\hline\hline
Survey & Expected Total 1-year & SNII Redshift  \\
Name & SNII Detections & Range \\
\hline
SDSS-II$^*$ & $1.70 \times 10^2$ & 0.03 $<z<$ 0.37  \\
DES & $2.74 \times 10^3$ & 0.06 $<z<$ 1.20  \\
Pan-STARRS & $5.14 \times 10^5$ & 0.01 $<z<$ 0.89  \\
LSST & $3.43 \times 10^5$ & 0.01 $<z<$ 0.89  \\
\hline\hline
\end{tabular}

Note:  {}$^*$Reflects SDSS-II supernova scan season of 3 months
  per calendar year.
\end{table}

For a given survey with a fixed scanning sky coverage $\oscan$,
we can determine the total number
of supernovae expected in each redshift bin.
We can also forecast the accuracy of the
resulting survey determination of the CSNR.
Namely, we can turn our sky coverage--monitoring time
result (Figure \ref{fig:dtdomega_r}) 
into a specific prediction for
the needed time to determine the CSNR to
a given precision. 
In practice, this amounts to a determination of the
redshift range over which different surveys
can measure the CSNR.
Our detailed predictions appear in Appendix \ref{sect:app-surveys};
here we summarize the results.

Several main lessons emerge from considerations of specific surveys.
When Pan-STARRS and LSST are online, these surveys will
collect a core-collapse supernova harvest far larger than 
the current set of events ever reported.
This alone will make synoptic surveys a transformational
point in the study of supernovae.

Moreover, synoptic surveys will detect core-collapse events
over a wide redshift ranges.
Table \ref{tab:sn} summarizes
the supernova redshift ranges correspond 
to the most likely $\mlim$ of the surveys.
The total supernova harvest depends sensitively on the survey depth,
and in Appendix \ref{sect:app-surveys}
the sensitivity to $\mlim$ is shown.
To determine the redshift ranges shown in Table \ref{tab:sn}, 
we set an (arbitrary) lower limit on the number of
total supernova counts at $N_{\rm min} = 10$.
We choose a lower redshift limit $z_{\rm min}$ such the 
cumulative survey supernova count in one year is 
$N_{\rm survey}(<z_{\rm min}) = N_{\rm min}$. 
Similarly, the upper limit $z_{\rm max}$ is set by
$N(>z_{\rm max}) = N_{\rm min}$ is the number of supernovae detected 
within redshift $z=z_{min}$ within a year. 

As seen in Tables \ref{tab:sn} and \ref{tab:redshift-r},
the future surveys will find abundant supernovae
over a wide redshift range.
At low redshifts, 
the surveys will detect nearly all of the supernovae
within the nearby cosmic volume accessible in their sky coverage. 
So surveys with a large $\oscan$, 
such as LSST and Pan-STARRS, have 
$z_{\rm min}$ which does not depends on $\mlim$. 
For DES, $\oscan$ is not as large, so that the number 
of supernovae brighter than $\mlim = \mag{21}$ does not accumulate 
to $N(<z_{\rm min})$=10 until $z_{\rm min}$=0.081. 
But for depths fainter than $\mlim = \mag{24}$, 
the survey does become volume-limited and the supernova
counts accumulate to 10 at the same redshift. 
The upper limit of the redshift $z_{\rm max}$
depends not only on sky coverage but also survey depth.
For planned survey depths, DES will gather core-collapse
supernovae to about
$z \simeq 1.20$; LSST will extend to 
$z\sim 0.89$, 
and could go further in modes with
smaller sky coverage but deeper exposure.

The large supernova counts and wide redshift ranges
together mean that surveys will, by direct counting,
map out the CSNR to high precision out to high redshifts.
Future surveys
should easily achieve 10$\%$ statistical precision for the
CSNR for redshifts around which the survey's counts peak.
We see that, with $\mlim = \mag{23}$,
LSST will reach out to $z \sim 0.89$, presuming that the
relative fraction of the brightest, farthest-reaching events (of Type IIn and Ic)
do not evolve with redshift.
If so, then by direct counting future survey should witness
the sharp CSNR rise. 
With deeper exposures and corresponding increases in redshift reach,
surveys could begin to test for the behavior of the CSNR
above $z=1$, a regime that is currently poorly understood.

Both the survey yields of supernova discoveries,
as well as their redshift ranges, are strong functions of
survey depth.
As shown in Table \ref{tab:redshift-r} 
of Appendix \ref{sect:app-surveys},
each magnitude increase in survey
depth yields a large enhancement (a factor $\sim 3$) in total supernova
counts.
This in turn leads to large enhancements in
redshift range, and thus in the range over which the CSNR is measured.
As shown in Appendix \ref{sect:app-surveys}, increased
monitoring time needed to achieve higher $\mlim$ 
will come at some cost, though this will be partially offset by the higher
supernova yield in a deeper exposure.
Finally, for the large population of low-redshift supernovae,
deeper surveys will lead to better lightcurve determination,
allow for a more accurate photometry over a larger 
brightness range and thus longer timescales.

As noted in \S \ref{sect:snlf},
our fiducial results are for supernova peak magnitudes
whose luminosity functions
(each of which is one or two gaussians for each core-collapse type)
are nonzero only within 
$|M - M_{\rm mean}| < 2.5 \sigma$ away from the mean.
This arbitrary cutoff is meant as a compromise which shows
the effect of nonzero width of the luminosity functions, 
without extrapolating too far into the tails in which there
is as yet no data.
To give a feel for the sensitivity of our results to the
assumed luminosity function width
we repeated our analysis for  luminosity functions
with larger and narrower $|M - M_{\rm mean}|$ ranges,
(but with fixed observed intrinsic $\sigma$).
We find that the total supernova counts vary less than 0.88\%
when $|M - M_{\rm mean}|$ ranges from $2\sigma$ to $3\sigma$;
this insensitivity reflects the fact that the bulk of supernova counts
are from events near the means of the distribution.
One the other hand, we found that the maximum observed supernova redshift 
(and thus the depth to which one can probe the CSNR)
is very sensitive to the choice of $|M - M_{\rm mean}|$.
For example, the LSST maximum supernova redshift in 1 year is
$z_{\rm sn,max} = 0.89$ for our fiducial choice of 
$|M - M_{\rm mean}| = 2.5 \sigma$, as seen in Table \ref{tab:sn}.
On the other hand for $|M - M_{\rm mean}| = 2\sigma$ and $3\sigma$,
we find $z_{\rm sn,max} = 0.73$ and 1.06, respectively.
Here rare, intrinsically bright events determine the redshift reach,
and the deeper the luminosity function reaches into the bright-end tail,
the larger the resulting $z_{\rm sn,max}$.
Thus we would expect the intrinsically brightest events,
of Type Ibc and Type IIn, to give the greatest redshift reach.
Indeed, \citet{cooke} has recently illustrated how
Type IIn events can be mapped out to $z>2$ by ground-based
8 meter-class telescopes.

Of course, all of our forecasts assume that the luminosity functions
of each supernova type, and the relative frequencies among
the supernova types, all remain unchanged at earlier epochs.
However, it is entirely plausible and even likely that these
properties could evolve, e.g., with metallicity and/or environment.
These effects are likely to be crucial in determining the true
redshift reach of future sky surveys, and thus predictions such as ours
will improve only as real supernova data becomes available 
with good statistics at ever-increasing redshifts, and one
can directly constrain and/or measure evolutionary effects.
Moreover, the relatively small sample sizes available to
\citet{richardson}
could well lead to underestimates of the true range of
luminosities of each type.  For example,
\citep{gezari} very recently report 
of an unusually bright
Type II-L event, SN 2008es, with peak magnitude $M_B \simeq -22.2$, 
far outside of the 
absolute magnitude range we have adopted for this subtype.

Indeed, the enormous statistics gathered by future surveys
will allow for cross-checks and empirical determination of 
other evolutionary and systematic effects.  
A major such effect is dust obscuration, to which we now turn.

\section{Dust Obscuration:  Disentangling the Degeneracies
and Probing High-Redshift Star-Forming Environments}

\label{sect:dustfuture}

The loss of some supernovae due to dust obscuration
must be understood accurately and quantitatively 
in order to take full advantage of the large
data samples of supernovae which {\em are} detected.
As noted above in \S \ref{sect:dust},
currently we have very limited knowledge of 
supenova extinction and particularly its evolution,
and most of what is reliably known is based on empirical studies
of supernova counts.
Precisely for this reason,
future surveys offer an opportunity to address this problem
in great detail by leveraging the enormous numbers of supernovae
of all types, seen 
over a wide range of redshifts
and in a wide range of environments.
Here we sketch a procedure for 
recovering this information.

Future surveys will produce well-populated
distributions of supernovae; these encode
information about extinction and reddening due to dust.
Specifically, in redshift bin $\Delta z$ around $z$
one can measure, often with very high statistical accuracy,
the apparent magnitude distribution for each subtype of core-collapse events.
These distributions can be made for all bands, but as we have
shown, detections and/or light curve information will be
most numerous in $r$ and $g$ bands; we will focus on these
for the purposes of discussion.
Within a redshift bin,
the distance modulus $\mu$ is fixed, and the light curve and associated
$K$-correction should reflect intrinsic variations within the
core-collapse subtype.

Thus, for a given core-collapse subtype
and redshift $z$, one can construct histograms of $r$ and $g$
peak magnitudes.
From redshift and supernova type, one can
compute the distance modulus and
$K$-correction,
and use these to infer, for each event, the
dust-obscured peak magnitude
$M_{\rm dust} \equiv m_{\rm obs} - \mu(z) - K(z) = M_{{\rm peak}} + A$
where $M_{\rm peak}$ is the intrinsic peak magnitude for
the event, and $A$
is the extinction for this event 
in its host galaxy.
By comparing two passbands we can also evaluate colors,
for example $g - r = M_{g,{\rm peak}} - M_{r,{\rm peak}} + E(g-r)$,
where $E(g-r)$ is the reddening.
In general, within a redshift bin
we expect the $A$ and $E$ to vary on an event-by-event basis,
reflecting the properties of dust along the particular line of sight
through the particular host galaxy.

Invaluable insight into these issues comes from
the \citet{hatano}
analysis of extinction in observation of local supernovae.
These authors argue that the data are consistent with
very strong dependence of extinction with the inclination of
the host galaxy; 
this alone guarantees that $A$ must
vary strongly from event to event even within subtypes.
\citet{hatano} also argue that the
variation of dust column with galactic radius
also suggests that extinction is responsible for
the paucity of supernovae at small radii. \citep{shaw}.
Finally, \citet{hatano} also point out 
that core-collapse events are
more extincted than Type Ia events because the Ia's have a higher
scale height and thus are more likely to occur in less extincted regions.

On an event-by-event basis,
intrinsic light curve and color evolution are
degenerate with dust evolution.
However, the large sample size may allow for
a physically motivated empirical approach to lifting
this degeneracy.
If on theoretical grounds we can assume that 
at least one core-collapse subtype has negligible
intrinsic evolution in its lightcurve,
then for that subtype $M$ and $K$ are effectively known and moreover are
constant across events in a particular redshift bin.
In this case, the apparent magnitude and color distributions
can be directly translated into distributions of
extinction and reddening.
By comparing these distributions (or e.g., their means and
variances) across different redshifts, one directly probes
dust evolution.


Moreover, if one can use one core-collapse subtype as an approximate
``standard distribution'' from which to extract dust properties, 
one might press further by assuming that other core-collapse
events will be born in similar environments and thus encounter
similar extinction and reddening.  
One can thus use the empirically determined dust evolution to
statistically infer the
degree of {\em intrinsic} lightcurve variation in 
the other core-collapse subtypes.

If subtype can be firmly established,
comparison of magnitude distributions
of different core-collapse subtypes
allows for a purely {\em empirical} approach.
namely, one can compare the evolution of the
magnitudes distributions of different subtypes.
One could first provisionally treat each supernova subtype as 
a ``standard distribution'' with no intrinsic evolution;
then for each subtype one would infer dust extinction and
reddening at each redshift.  It is reasonable to expect that the
different subtypes sample the same dust properties,
as long as the host environments are
not systematically different for the different subtypes (all of which
arise in massive-star-forming environments).
Indeed, \citet{nugent} have performed such an analysis to
use $V-I$ colors of Type II-P events to infer reddening
for events out to $z \sim 0.3$.

A comparison of the dust extinction inferred the different
subtypes amounts to a test for intrinsic variation.
With information from multiple subtypes, it may be possible
to isolate dust effects common to all, and intrinsic variation
peculiar to each subtype.
For example, if one subtype distribution evolves significantly more
than another (e.g., one subtype variance grows more than another)
then the difference in variance must be intrinsic, and
that the lesser variance is an upper limit to the variance
due to dust effects.

The ability to empirically measure extinction
depends on the intrinsic width of the $A(z)$ distribution,
and on surveys' ability to probe this distribution.
At low redshift, \citet{hatano} find a wide ($> \mag{1}$)
range of extinctions, much of which they attribute to
inclination which will remain an issue at higher redshift.
On the other hand, as a given survey pushes to higher redshift,
progressively less of the distribution is observable.
For the case of LSST, we see in Fig.~\ref{fig:LSST}
that with $\mlim = \mag{23}$ the least obscured events
are visible out to redshift $z \sim 1$, while
those which have suffered $A_r = \mag{1}$ of extinction
would correspond to the $\mlim = \mag{22}$ curves,
which reach to about $z \sim 0.5$.  Thus over this
shallower redshift range, extinction can be probed
in detail, but with a narrowing observable range at higher
redshift.

If future surveys can empirically determine effects of
dust evolution, this would not only remove a major
``nuisance parameter'' for supernova and cosmology
science, but also gain information of intrinsic interest.
Namely, we will learn about the cosmic distribution and evolution of
host environments of
supernovae and thus of star formation.

\section{Discussion}
\label{sect:discuss}

The large amount of core-collapse supernovae 
observed by synoptic surveys
will yield a wealth of data and
enormous science returns.
Here we sketch some of these.

\subsection{Survey Impact on the Cosmic Supernova
and Star Formation Histories}

As we have indicated in the previous section,
synoptic surveys will determine cosmic core-collapse supernova
rate with high precision out to high redshifts.
Moreover,
with the large number of supernova counts, 
and with light curves and host environments known,
the total cosmic core-collapse redshift history can be subdivided
according to environment and/or supernova type.
For example, with photometric data alone
one can determine to high accuracy 
correlations between supernova rate and host galaxy luminosity
and Hubble type.
One can compare supernova rates in field galaxies versus those
in galaxy groups and clusters.
Using galaxy morphology one can investigate correlations
between supernovae and galaxy mergers.
With the addition of 
spectroscopic information one can also search for correlations with
host galaxy metallicity.

The CSNR is also tightly related to the cosmic
star-formation rate. Therefore with the high precision 
CSNR, and assuming an unchanging initial mass function,
one can make a similarly precise measure of the 
cosmic star-formation rate.
On the other hand, one can test for environmental and/or redshift
variations in the initial
mass function, by comparing the supernova rates based on
direct survey counts with the star-formation rates
determined via UV and other proxies.

In addition to core-collapse explosions,
synoptic surveys will of course by design also
discover a similarly huge number of Type Ia supernovae. 
Thus the Type Ia supernova rates can be compared to
those for 
core collapse events.
As has been widely noted 
\citep[e.g.,][and references therein]{gal-yam,watanabe,oda,sb}
this will yield information about 
the distribution of time delays between the core-collapse
and thermonuclear events.
Moreover, one can explore differences in the
environmental correlations for the two supernova types (and subtypes).

\subsection{Survey Supernovae as Distance Indicators:  
the Expanding Photosphere Method}

Type Ia supernovae have become the premier tool for
distance determinations at cosmological scales, thanks to
their regular light curves, high peak brightnesses, and 
relatively less dusty environments.
Nevertheless, given the importance of the cosmic distance
scale, and the 
ongoing need for systematic crosschecks and calibration,
it is worthwhile to consider other methods.
Core-collapse events offer just such a method, via 
the expanding photosphere/expanding atmosphere method.

This method was originally conceived by \citet{baade} and \cite{wesselink}
for study of Cephieds;
\citet{kk} applied the Baade-Wesselink method to supernovae.
The key to the technique is to exploit
the simple kinematics of a newborn supernova remnant:
the freely-expanding photosphere grows in size as
$R = vt$.  Thus, for purely blackbody emission,
the luminosity grows with size (i.e., time) as $L = 4\pi R^2 \sigma T^4$.
With good sampling to measurements time $t$ since explosion,
and spectroscopic inference of $v$ and $T$, 
one can recover the luminosity.  In principle, therefore,
one can use the
explosion as a standard candle.

In practice, this method has been slow to mature.
Until recently,  the agreement with independent 
distance measures has been only good to within a factor $\sim 2$
\citep[e.g.,][]{vt}.
The complex (out of local thermodynamic equilibrium) spectra of
supernovae has proved difficult to adequately model.
However, recently important advances have been made in
the  radiation transfer modeling of 
young supernova remnants
and its fitting to spectra of local supernovae
\citep{baron,dh,dh07}.
Because of this, the expanding photosphere (or more properly,
expanding atmosphere)
method now appears to be reaching consistency with other 
distance measures;
this method
now shows agreement approaching the $\sim 10\%$ level.
Similar pecision now seems possiblue using 
a separate, empirical method  \citep{hp}
which exploits the observed correlation between luminosity and
expansion velocity of Type II-P events.
This opens up core-collapse supernovae as alternative distance indicators.
Indeed, several group \citep{nugent,poznanski,olivares}
have already applied this method to various 
collections of Type II-P observations,
yielding tight Hubble diagrams out to $z \sim 0.3$.

To use this method as it is currently envisioned, follow-up
spectroscopy is mandatory for {\em each} 
event (see \S \ref{sect:followup}),
with photometric surveys identifying the candidates.
Obviously, for the largest surveys, in practice
only a tiny fraction of core-collapse events could be
studied in a (separate) spectroscopic campaign, particularly
given that the most common core-collapse types are
intrinsically dimmer than Type Ia events and thus
require longer exposures to obtain spectra.
Followup requirements thus are the limiting factor for
the use of core-collapse events as distance indicators.

The situation for Type Ia supernovae is better-studied and
also potentially more hopeful.
Recent work \citep{poznanski,kim,kunz,blondin,wang2007,kuznetsova,sako}
suggests that photometric redshifts of Type Ia events
near maximum light could be obtained with sufficient
precision (give a low-redshift training set)
to provide useful dark energy constraints without spectroscopy.
Whether photometric-based distances can be
derived for core-collapse events with sufficient accuracy
remains to be seen.  It nevertheless seems
to us a worthy object of further study.
In this context it is worth noting that DES plans to do followup
spectroscopy on $\sim 25\%$ of Type Ia events \citep{des}.
We suggest that at least some modest fraction of this follow-up time
be dedicated to core-collapse monitoring.

\subsection{Other Science with Cosmic Supernovae} 
 
The physics, astrophysics, and cosmology of cosmic core-collapse supernovae 
is a fertile topic; with our detectability  
study in hand, a wide variety of problems present themselves. 
Here we sketch these out; we intend to return to these
in future publications.

The huge harvest of core-collapse events will open new windows 
onto other aspects of supernova physics. 
For example, the physics of black hole formation in supernovae, 
and the neutron-star/black hole divide, remain important open 
questions.  \citet{bs}
have estimated the rates of events with observable signatures 
of black hole formation; LSST should provide a fertile testing 
ground for these predictions. 
 
The elaboration of the cosmic history and 
specific {\rm sites} of high-redshift supernovae 
will also offer unique new information 
about supernova ``ecology'' -- i.e., feedback 
and cycling of energy, mass, and metals into the surrounding 
environment.  
For example, large surveys will offer the 
opportunity to study supernova rates 
as a function of host galaxy and galaxy clustering, 
shedding new light onto large-scale star formation and its 
connection with galaxy evolution. 
Moreover, DES and other surveys will discover an 
enormous number of rich galaxy clusters; the 
occurrence of both Type Ia and core-collapse events in clusters 
will offer important new insight into the origin 
of the very high metallicity of intracluster gas 
\citep{mg,maoz}.

Core-collapse supernovae also are the sources,
directly or indirectly, of high-energy radiation of
various kinds.  For example, supernovae act as accelerators of 
cosmic rays.  These in turn interact with 
interstellar matter  to produce high-energy $\gamma$-rays. 
\citet{pf} used then-available 
estimates of the cosmic star-formation rate 
to show that this $\gamma$-ray signal has 
a characteristic feature, and makes a significant 
part of the extragalactic $\gamma$-ray background 
around $\sim 1$ GeV.  With the successful launch of the 
high-energy $\gamma$-ray observatory GLAST, 
this component of the $\gamma$-ray background 
may for the first time be clearly identified. 
Regardlessly, a sharper knowledge of the cosmic supernova 
rate (and thus cosmic-ray injection rate) 
will work in concert with GLAST observations 
to probe the history of cosmic rays throughout the universe.

\subsection{Comparison with Type Ia Survey Requirements} 

The characteristic of Type Ia supernovae are 
in general very similar to the core-collapse supernovae. 
Hence the observational requirements for their identification
in synoptic surveys are also very similar. 
The rest-frame, full-width at half-maximum timescale 
for Type Ia supernovae is $\sim 20$ days.
Therefore surveys will need a scan cadence of a few 
days in order to get a well-sampled light curve.
For example, the Pan-STARRS strategy for Type Ia discovery
is to sample the light curve 
every 4 days \citep{tonry,pan-starrs-sne}. 
As we have discussed, 
this sampling frequency is also suitable for the core-collapse
supernovae which the time scale of the light curve also 
last a few weeks.

\subsection{Redshifts and Typing from Photometry and Followup Spectroscopy}
\label{sect:followup}

Survey supernovae become scientifically useful only when
one can  establish, at the very least,
their redshift and whether they are core collapse or Type Ia.  
Since followup spectroscopy will not be possible
for the large numbers of future events, photometric redshifts
will be needed.  For events in which a host galaxy is 
clearly visible, one can use photometric redshifts of the
hosts.  Here one is helped by the ability of surveys to
stack all of the many (non-supernova) exposures to obtain
a much deeper image than those with the supernovae.
Once the host redshift is know, the supernova type
must be determined.  
Baysean analysis techniques and software
\citep{dg,poznanski}
have been developed
to distinguish
both Type Ia and
core-collapse events.
These authors find that type discrimination 
depends crucially on the accuracy with which the redshift is known.
For spectroscopic redshifts, their methods is extremely accurate,
and for photometric redshifts the method is still quite good,
though in this case misclassifications can reach $15-25\%$ depending
on $\sigma_{{\rm photo}-z}$.

Followup spectroscopy on a subset of events
will be essential to calibrate the
accuracy of the photometric typing (and host redshifts).
In particular, spectroscopy will be invaluable in identifying and
quantifying catastrophic failures in the typing algorithms;
on the basis of these it may be possible to refine the
routines.
As noted in the previous section, followup is also required
for any events one hopes to use in distance determinations.

For events without clear host galaxies and without followup, one must
resort to photometric redshifts and typing based on the supernova
light curve itself, in whatever bands are available.
It is not clear that this can be done with any
reliability on 
an event-by-event basis
As \citet{poznanski} emphasize, 
one might make statistical statements about the
types and redshifts of the entire class of ``hostless'' events.
Here spectroscopic followup will be essential, not only
for determining the supernova redshift but also the nature of
the underlying host.

\section{Conclusions and Recommendations for Synopic Surveys}  
\label{sect:conclude}

The next ten years will witness a revolution in our
observational knowledge of core-collapse supernovae.
Synoptic sky surveys will reap an enormous harvest of
these events, with tens of thousands discovered
in the near future, culminating with of order
100,000 seen annually by LSST.
These data will reveal the 
supernova distribution in space and time over much of cosmic history.
The needed observations are naturally a part of the
scanning nature of these
surveys, and require only that
core-collapse events be included in the data analysis pipeline.

The potential science impact of this unprecedented supernova sample
is enormous.
We have discussed ways in which the photometric supernova data
alone will contribute in significant and unique ways 
to cosmology and astroparticle physics, as well as to
studies of core-collapse and supernova evolution themselves.
We illustrate one such application by demonstrating 
how to recover the cosmic supernova rate from 
the redshift distribution of supernova counts in synoptic
surveys. 
The large datasets ensure that the
statistical error will be very low,
and the first large surveys will
rapidly determine the CSNR to precision
exceeding that of current data based on observation of
massive-star proxies.

With the addition of spectroscopic followup 
observations, the survey-identified core-collapse supernovae
can be used as distance indicators.  
Thanks to recent advances in the 
phenomenology of supernova spectra and the
modelling of their expanding atmospheres,
the early light curves provide standardizable candles.
This expanding photosphere/atmosphere method
could provide a cross-check for the cosmic distance
scale as inferred from Type Ia supernovae.

To summarize our recommendations for synoptic surveys,
in order to capitalize on this potential:
\begin{enumerate}

\item
Include core-collapse supernovae (all Type II as well as
Types Ib and Ic) in the data analysis pipeline.

\item
Include a scanning mode in which the depth $\mlim$
is as large as possible, in order to maximize
the supenova redshift range. Surveys which modes which probe 
down to $\mlim= \mag{26}$
could discover many supernovae (both core-collapse and Type Ia)
approaching redshift $z \sim 2$.

\item
Adopt scanning cadence of revisits every $\sim 4$ days,
in order to capture core-collapse events at peak brightness, and
to obtain a well-sampled lightcurve.
This timescale also appropriate for Type Ia events.

\item
Allocate some followup spectroscopy to core-collapse events.
This will calibrate photometric Type Ia/core-collapse typing
and typing among core-collapse subtypes, and will be
particularly crucial for probing the nature of
events in which a host galaxy is not seen.

\end{enumerate}

We close by re-emphasizing that these recommendations
require only modest efforts in analysis, little to no modification
of the strategies already in place for Type Ia searches,
and some commitment of followup spectroscopy.
Thus a small extra investment of resources
will reap handsome scientific rewards
as we open our eyes to the incessant rise and fall of
these beacons marking
massive star death throughout the cosmos.

\acknowledgments 
It is a pleasure to
thank Joe Mohr, Vasiliki Pavlidou, Tijana Prodanovic, 
Jon Thaler, Yun Wang, 
David Weinberg, and Michael Wood-Vasey for valuable discussions. 
We are particularly indebtted to
Josh Frieman, Avishay Gal-Yam, John Beacom, and David Branch for
detailed constructive feedback on an earlier draft, 
which has greatly improved this paper.

\bigskip

{\em Note:} In the final stages of writing this paper
we became aware of the work of \citet{young}.
These authors discuss core collapse rates in sky surveys,
with a focus on events in low-metallicity environments.
Where it is possible to compare and when we adopt their parameters
(particularly $\mlim$), our analyses seem to be
in broad agreement.

\appendix

\section{Appendix:  The Supernova/Star-Formation Connection}
\label{sect:app-csfr}

The star-formation rate and supernova rate for any astrophysical site
are intimately related.  Moreover, in many applications such as ours
the timescales of interest are much longer than the $\sim few$ Myr
supernova progenitor lifetimes.
In this case, the star-formation rate and supernova rates are
proportional.  This is expressed above in eq.~(\ref{eq:snr-sfr}).
The constant of proportionality can be obtained from the
initial mass function $\xi(m)$.
In stellar mass range $(m,m+dm)$ the number of
new stars is proportional to $\xi(m) \, dm$, while
the mass of new stars is $m \, \xi(m) \, dm$.
Thus the number of supernovae per unit new star mass--i.e.,
the conversion between star-formation and supernova rates--is
\beq
\frac{\csnr}{\csfr}
 = \frac{\int_{\rm SN} \xi(m) \ dm}{\int  m \ \xi(m) \ dm}
 = \frac{\int_{\rm SN} \xi(m) \ dm}{\int_{\rm SN}  m \ \xi(m) \ dm}
   \frac{\int_{\rm SN} m \ \xi(m) \ dm}{\int  m \ \xi(m) \ dm}
 = \frac{X_{\rm SN}}{\avg{m}_{\rm SN}}
\eeq
where $X_{\rm SN} = \int_{\rm SN} m \, \xi \, dm/\int m \, \xi \, dm$
is the mass fraction of new stars that will go into supernovae,
and 
$\avg{m}_{\rm SN}  = \int_{\rm SN}  m \, \xi(m) \, dm/\int_{\rm SN} \xi(m) \, dm$ is the mean supernova progenitor mass.
For illustration, consider a Salpeter IMF $\xi(m) \propto m^{-2.35}$
over mass range $(0.5\msol,100\msol)$
and $\xi(m) \propto m^{-1.5}$ for the low mass range $(0.1\msol,0.5\msol)$
\citep[][their ``Salpeter A'' mass function]{bg}.
We also take supernova progenitors to lie
in the mass range $m_{\rm SN} \in(8\msol,50\msol$).
This gives $X_{\rm SN} = 0.15$,$\avg{m}_{\rm SN} \approx 15.95 \msol$,
and thus a star-formation/supernova conversion factor
$X_{\rm SN}/\avg{m}_{\rm SN} = 0.00914 \msol^{-1}$.
The uncertainty here is significant, probably
about a factor of 2.

The cosmic star-formation rate can be estimated from
a number of observables tied to massive (i.e.,
short-lived thus ``instantaneous'') star-formation.
Proxies often adopted are
the UV
and/or H$\alpha$ luminosity densities \citep{madau}.
Of these, UV light has a more direct connection with
massive stars, but is also affected more by the dust
extinction than the H$\alpha$ light \citep{strigari05}.
Most cosmic star-formation
studies find a sharp increase in the rate
up to $z=1$, but there remains a large
uncertainty of the star formation rate
at higher redshift.
In this paper, we adopt the
\citet{cole} fitting formula for the cosmic star-formation rate
\beq
\csfr^{\rm cole} =  \frac{a+bz}{1+(z/c)^d} \ h \, M_{\odot} \, \rm yr^{-1} \, Mpc^{-3}
\eeq
where$(a,b,c,d)=(0.017,0.13,3.3,5.3)$, which are one of the
best-fitted parameters according to current
observing data found by
\citet{hopkins}.
Using the conversion factor of \citet{hopkins} and our adopted Hubble constant,
this gives
a local rate of
$\csnr(0) = 1.1 \times 10^{-4} \rm \ SNII \, Mpc^{-3} \, yr^{-1}$.
The benchmark CC SNe rate
rises to a peak around $z = 2.5$ and then slowly declines
at high redshift.

To illustrate the impact of different star formation,
we also did all calculations with an alternative CSNR.
Here we normalize to the current observed, counting-based
CSNR  \citep{botticella}, and take the shape from
the a fitting function 
of the \citet{hippelein} star formation rate.
We also lower the rate by 30\% because 
we want the alternative CSNR to be as much different 
as the benchmark CSNR as possible and \citet{hopkins} 
suggested that the uncertainty of the normalization 
of the star formation rate is about 30\%.
The differences between this and our fiducial rate
gives a sense of the current rough but not perfect agreement
between the CSNR as inferred indirectly from progenitor light
(sometimes reprocessed)
and directly from counting.

\section{Appendix: Supernova Predictions for Upcoming Synoptic Surveys}
\label{sect:app-surveys}

\begin{figure}[!hb]
\includegraphics[width=0.8\textwidth]{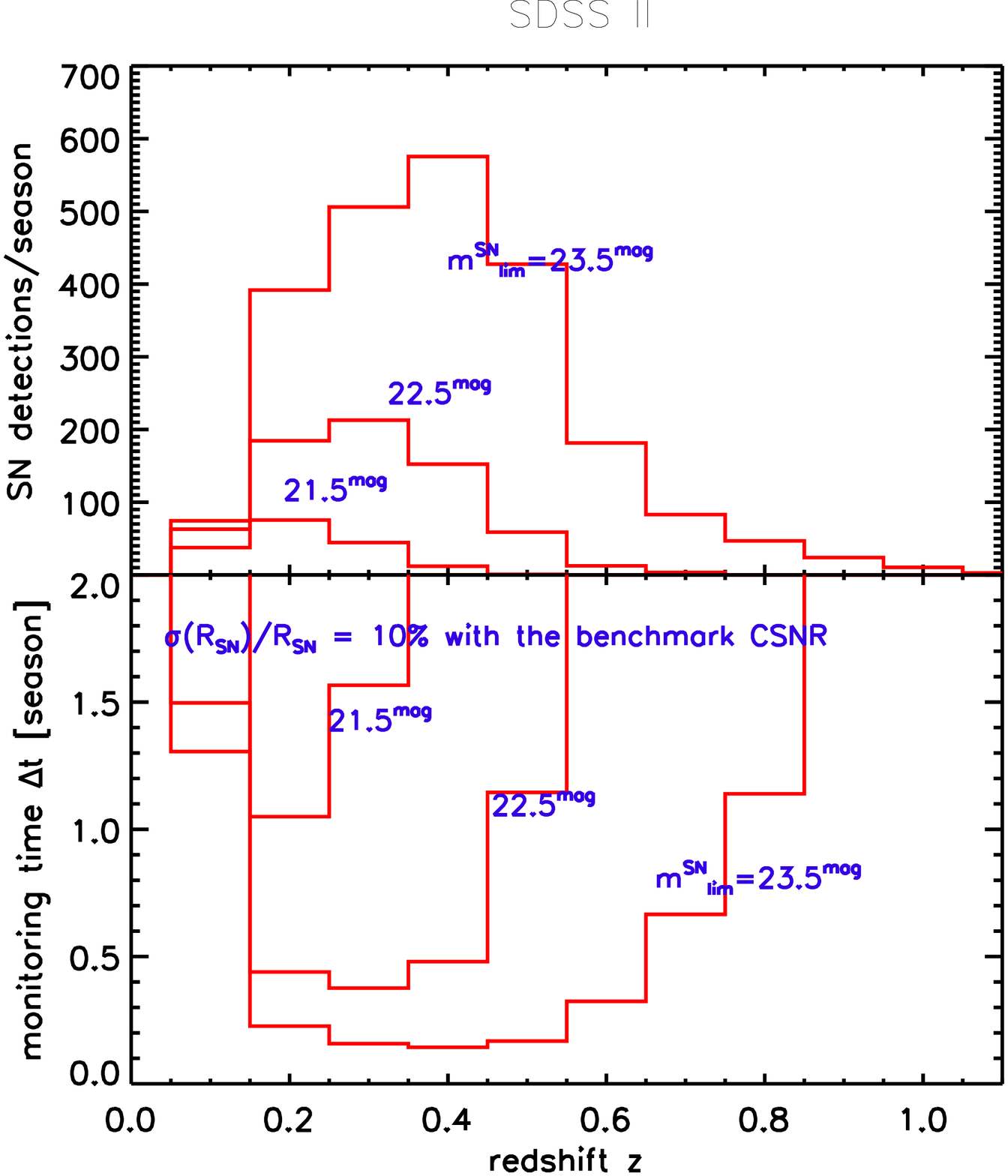}
\caption{Sloan Digital Sky Survey (SDSS-II).
All the results plotted here are using the benchmark CSNR.
{\em Top panel:} The number of supernovae observed in one 
scan season of 3 months per year,
in redshift bins of with $\Delta z = 0.1$.
Results are shown for a fixed scan sky coverage
$\oscan = 300 \ {\rm deg}^2$,
and the survey depth as labeled.
{\em Bottem panel:} The monitoring time $\Delta t$ needed 
in order to determine the cosmic supernova rate to a $10\%$ precision. 
}
\label{fig:SDSS}
\end{figure}  

\begin{figure}[!hb]
\includegraphics[width=0.8\textwidth]{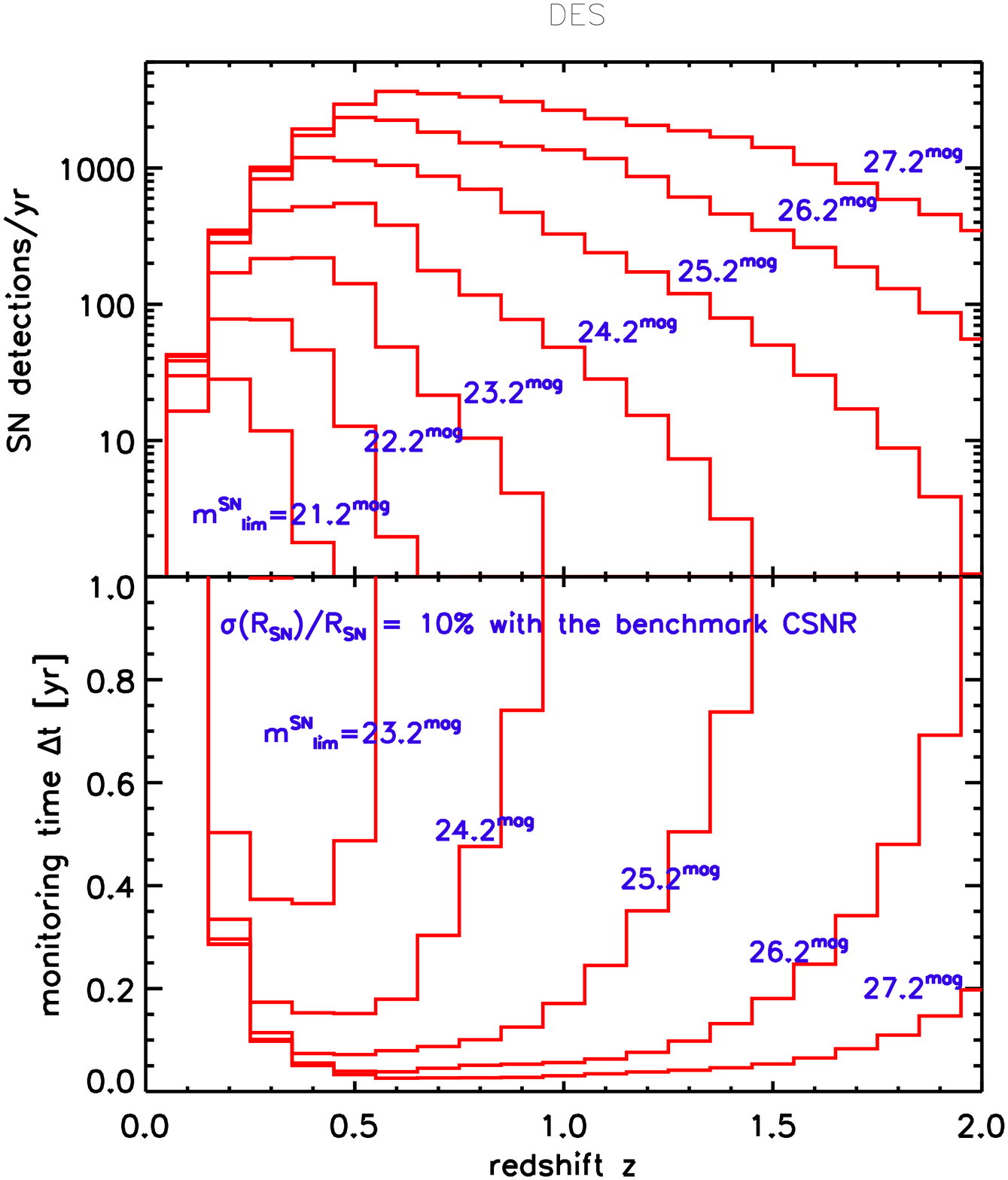}
\caption{
As in Figure \ref{fig:SDSS}, for the Dark Energy Survey (DES). 
Here results are show for different $\mlim$, but for a fixed
scan sky coverage $\oscan = 40 \ {\rm deg}^2$.}
\label{fig:DES}
\end{figure} 

\begin{figure}[!hb]
\includegraphics[width=0.8\textwidth]{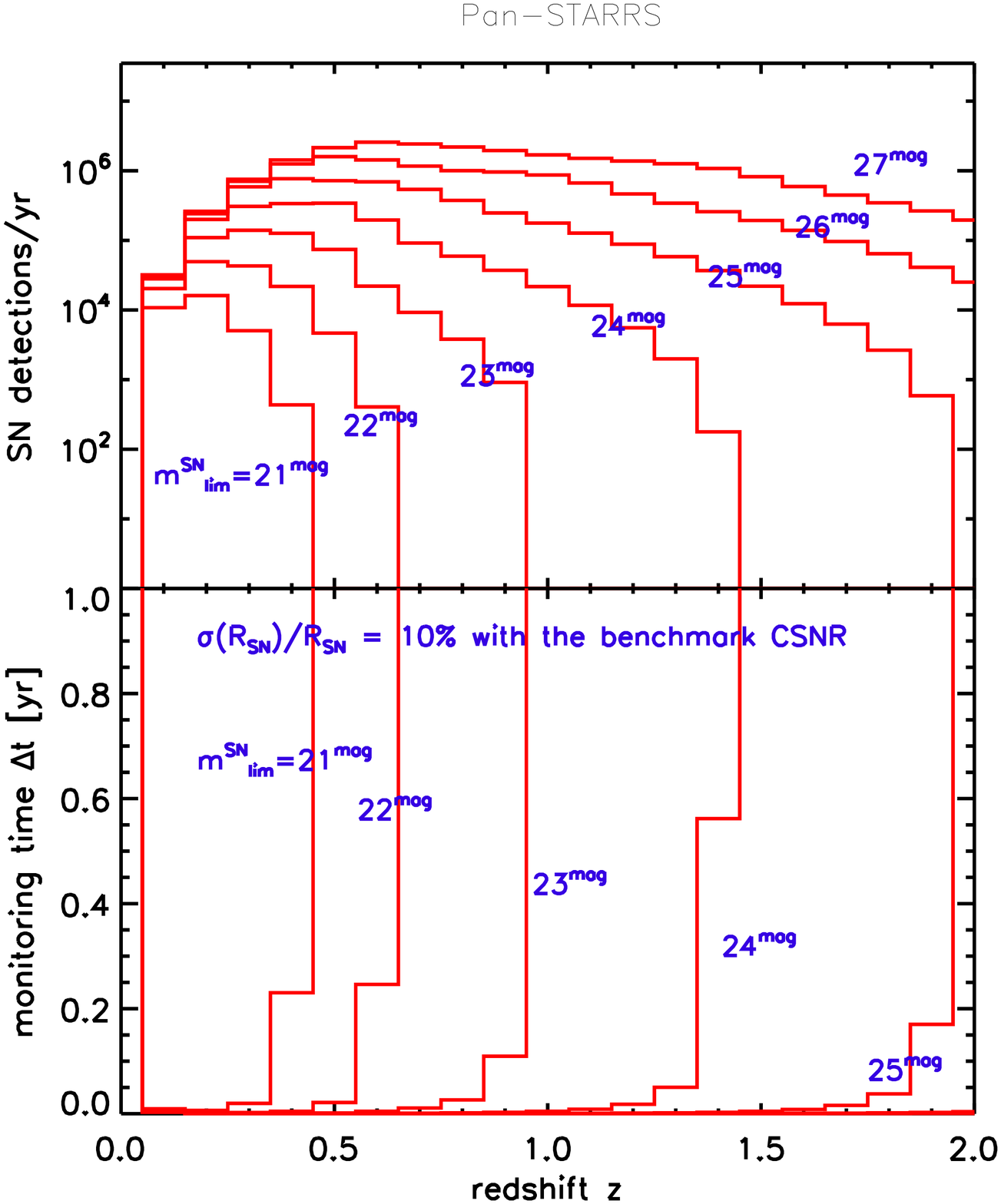}
\caption{
As in Figure \ref{fig:DES}, for Pan-STARRS.
We hold fixed $\oscan = 30000 \ {\rm deg}^2$.
}
\label{fig:Pan-STARRS}
\end{figure}

\begin{figure}[!hb]
\includegraphics[width=0.8\textwidth]{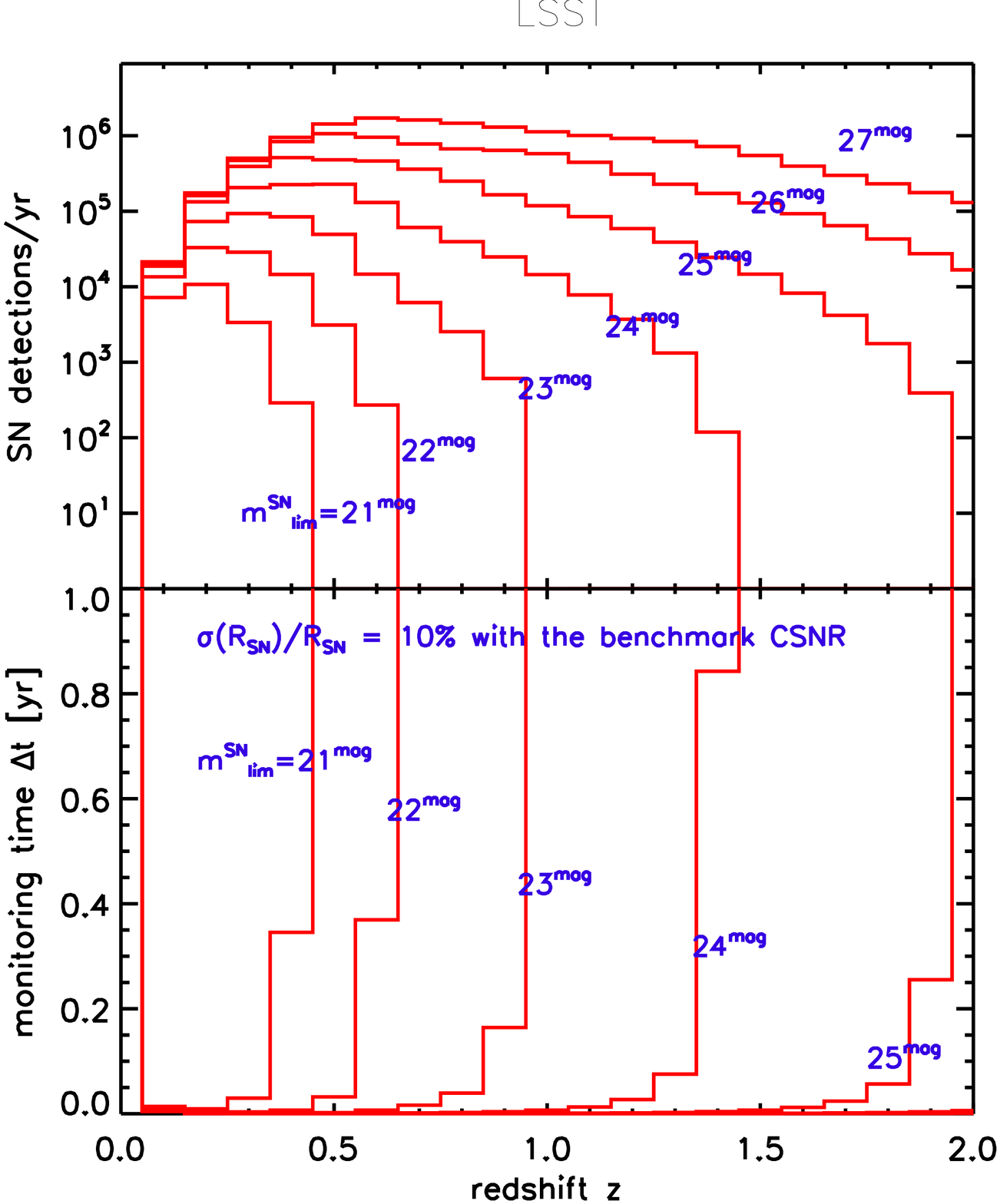}
\caption{
As in Figure \ref{fig:DES}, for Large Synoptic Survey Telescope
(LSST).
We hold fixed $\oscan = 20000 \ {\rm deg}^2$.
}
\label{fig:LSST}
\end{figure} 

\begin{table}[htb]
\caption{\label{tab:redshift-r}
Sensitivity to survey depth in $r$ band, for fixed sky coverage.}
\begin{tabular}{c|cc|cc|cc}
\hline\hline
$\mlim$  & \multicolumn{2}{c|}{DES} & \multicolumn{2}{c|}{Pan-STARRS} & \multicolumn{2}{c}{LSST}  \\
{[mag]} & Total SNe & Redshifts & Total SNe & Redshifts & Total SNe & Redshifts \\
\hline
21 & $4.32 \times 10^1$ & 0.081 $<z<$ 0.18
   & $3.24 \times 10^4$ & 0.007 $<z<$ 0.36
   & $2.16 \times 10^4$ & 0.008 $<z<$ 0.36 \\
22 & $1.86 \times 10^2$ & 0.066 $<z<$ 0.38
   & $1.40 \times 10^5$ & 0.007 $<z<$ 0.56
   & $9.32 \times 10^4$ & 0.008 $<z<$ 0.56 \\
23 & $6.86 \times 10^2$ & 0.064 $<z<$ 0.66
   & $5.14 \times 10^5$ & 0.007 $<z<$ 0.89
   & $3.43 \times 10^5$ & 0.008 $<z<$ 0.89 \\
24 & $2.19 \times 10^3$ & 0.063 $<z<$ 1.10
   & $1.64 \times 10^6$ & 0.007 $<z<$ 1.34
   & $1.10 \times 10^6$ & 0.008 $<z<$ 1.33 \\
25 & $6.32 \times 10^3$ & 0.063 $<z<$ 1.62
   & $4.74 \times 10^6$ & 0.007 $<z<$ 1.88
   & $3.16 \times 10^6$ & 0.008 $<z<$ 1.88 \\
26 & $1.55 \times 10^4$ & 0.063 $<z<$ 2.17
   & $1.16 \times 10^7$ & 0.007 $<z<$ 2.48
   & $7.73 \times 10^6$ & 0.008 $<z<$ 2.47 \\
27 & $3.17 \times 10^4$ & 0.063 $<z<$ 2.68
   & $2.38 \times 10^7$ & 0.007 $<z<$ 3.08
   & $1.58 \times 10^7$ & 0.008 $<z<$ 3.08 \\
\hline\hline
\end{tabular}
\end{table}

Figures \ref{fig:SDSS}--\ref{fig:LSST} 
show the supernova forecasts for
different surveys.  
For each we show the expected annual supernova harvest
$\Delta N_{\rm obs}(z)$ as a function of redshift,
with the sky coverage held fixed to the values
in Table \ref{tab:vitalstats}.
The count distribution across redshift bins are directly proportional
to the differential supernova rate distribution
$\Gamma_{\rm SN,obs}(z)$, via eq.~\pref{eq:snnumber}.
Thus the shapes the curves
follow those of 
$\Gamma_{\rm SN,obs}(z)$ as seen in
Figure \ref{fig:dNdodtdz_band}
an explained in the accompanying discussion.

Figures \ref{fig:SDSS}--\ref{fig:LSST} 
also show the survey scan time required for these data
to constrain the 
cosmic star-formation rate
in each redshift bin to within a statistical precision 
$\sigma(\csnr)/\csnr = 10\%$. 
We have seen (eq.~\ref{eq:csnr-sig})
that the precision at each bin is inverse
with the counts, $\sigma(\csnr)/\csnr = 1/\sqrt{\Delta N_{\rm obs}(z)}$.
Thus these panels show trends in which 
monitoring time decreases with the
counts per bin.
This mirrors the behavior 
shown in 
Figure \ref{fig:dtdomega_r}
and explained in the surrounding discussion.

Table \ref{tab:redshift-r} shows the effect of
survey limiting magnitude on redshift range
and total supernova harvest in the $r$-band. 
We see that each unit increase $\Delta \mlim = \mag{1}$ 
in survey depth yields a large enhancement,
in the total supernovae seen.
The supernova numbers
increase by a factor 4.3 when going from $\mlim = \mag{21}$ to
$\mag{22}$ to factor of 2.0 when going from 
$\mag{26}$  to $\mag{27}$.
Of course, there is a tradeoff in the needed exposure.
For the faintest objects at the highest redshifts, 
background noise dominates, and monitoring time
grows by a factor $(10^{2/5})^2 \sim 6.3$ per magnitude. 
Thus, including a narrower but deeper survey
mode will likely yield fewer supernovae,
but if judiciously implemented, this tradeoff may be worth the additional
redshift coverage.

\section{The $K$-Correction}
\label{sect:K-correction}

\begin{figure}[!htb]
\includegraphics[width=0.9\textwidth]{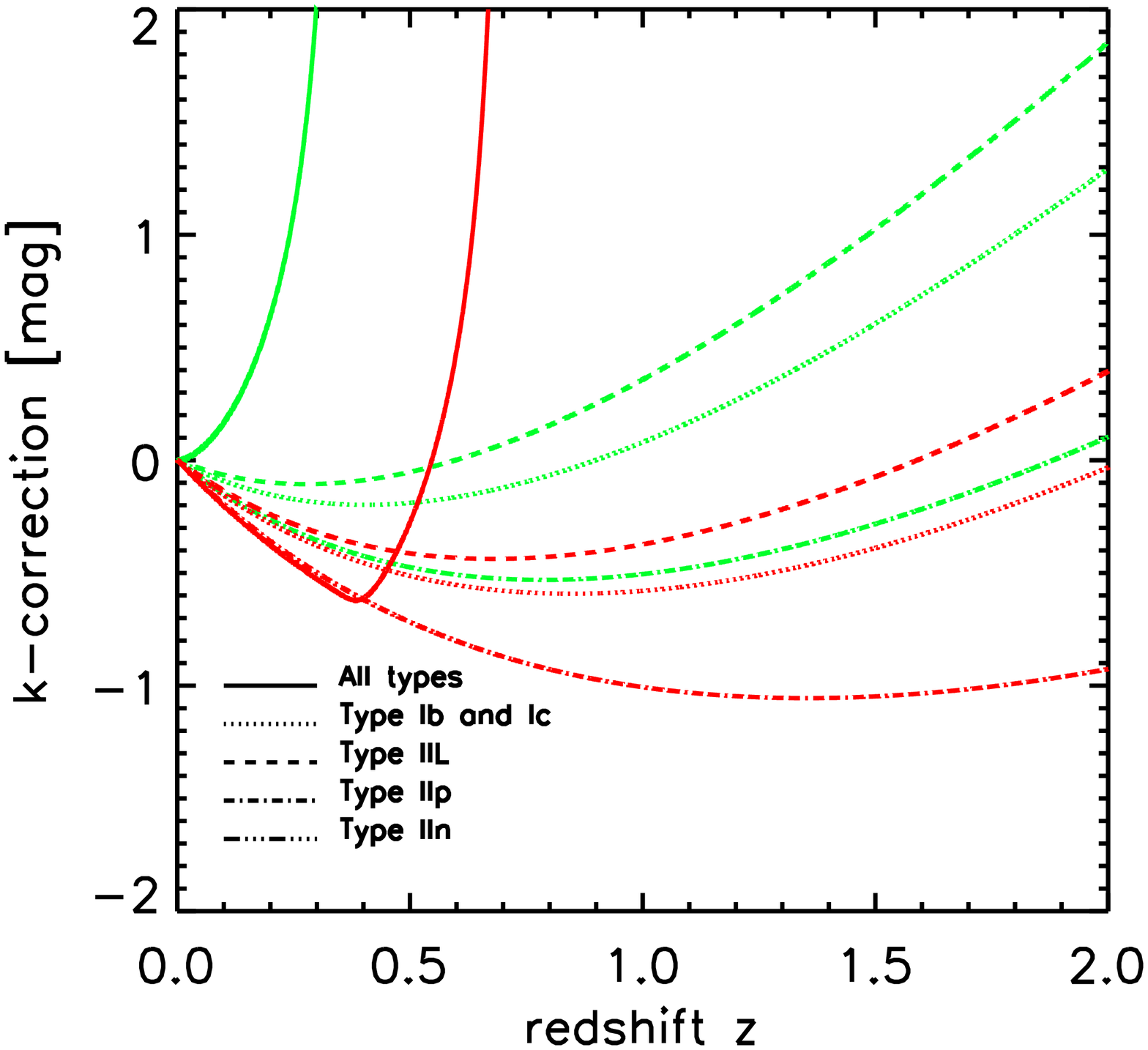}
\caption{$K$-correction of the four types of supernovae(I-bc,II-L,II-P,II-N) 
in $g$ and $r$ bands (green lines for $g$-band and 
red lines for $r$-band). 
The sharp upturn in the Ibc correction reflects a cutoff in
the supernova spectrum due to UV line blanketing; see discussion
in text.
\label{fig:kcorrection}
}
\end{figure}

The $K$-correction accounts for redshifting of the supernova spectrum
across the passbands in the observer frame.
In the context of the
Southern inTermediate Redshift ESO Supernova Search,
the elegant and instructive analysis of
\citet{botticella}
determine $K$-corrections for the $\sim 90$ supernova
confirmed and candidate events in their survey.
They found that the corrections depend strongly on
redshift, light curve phase, and on waveband.
In particular, the shifts from observed $V$ and $R$ bands
to rest-frame $B$-band
both typically have $K<0$, i.e, a negative correction,
particularly at early times most relevant here; this reflects
the blue colors of the early phases.
The corrections are at early times (within the first three weeks)
usually a shift  $|K| \la \mag{1}$, with the largest corrections
$K \sim \mag{-2}$.
As seen in eq.~\pref{eq:distmod},
a negative correction would reduce the apparent magnitude
and thus {\em improve} the observability
of the supernovae.

In this paper we compute $K$-corrections following the formalism of
\citet{kgp} in a single band, and add the color correction 
term (which described as $\eta_{xB}$ in Section \ref{sect:snmlim}) to 
make the corresponding correction when transfer 
from one band to another.
In band $x$, the $K$-correction is
\beq
K_x = 2.5 \, \log(1+z) \ + \ 
    2.5 \log \frac{\int F(\lambda) S_x(\lambda)d\lambda}
                  {\int F(\lambda /(1+z))S_x(\lambda)d\lambda}
\eeq
where $F(\lambda)$ is the unobscured, rest-frame spectral distribution 
of the supernova, and $S_x(\lambda)$ is the sensitivity of filter $x$.
We included five types of core collapse supernovae: 
Ib, Ic, II-L, II-P, IIn.
For each spectral type, we adopt a rest-frame spectrum $F(\lambda)$
following the prescription of \citet{dahlen},
who adopt blackbody spectra (sometimes slightly modified)
with different temperatures and time evolution.
We picked the temperatures 
which last for about a weak around the peak luminosity 
and treat them as a constant. Since the surveys we 
are interested in will have cadence less than 
a week, this should be a reasonable simplification.
For Type Ib and Ic, we choose a 15,000 K blackbody 
with cutoff at $\lambda < 4000 \AA$ because of 
UV blanketing.
We choose 11,000 K for Type II-L, 10, 000 K for Type II-P, 
and 14,000 K for Type IIn. 

Our resulting $K$-correction appear in Fig. \ref{fig:kcorrection},
plotted for the $g$ and $r$ bands which we will see are 
the optimal for supernova discovery.
The huge turnoff of the 
$K$-correction in Type I-bc is due to the short-wavelength 
cutoff in its spectrum.
A large value is set so that 
there will be no Type I-bc supernovae observed beyond the cutoff point. 
Regardless of the turnoff of Type I-bc, for low redshifts
the $K$-correction is 
in the range of $\mag{+2}$ to $\mag{-1.1}$ but typically is negative, 
which is in consistent with \citet{botticella}.

\newpage



\end{document}